\documentclass{article}
\usepackage{geometry}
\usepackage{amsmath}
\usepackage{amsfonts}
\usepackage{amssymb}
\usepackage{braket}
\usepackage[shortlabels]{enumitem}
\usepackage{xcolor}
\usepackage{graphicx}
\usepackage[section]{placeins}
\usepackage[sort&compress,numbers]{natbib}
\title{Revisiting the replica trick: Competition between spin glass and conventional order}

\author{Christopher L. Baldwin\textsuperscript{1} and Brian Swingle\textsuperscript{2}}
\date{{\small \textsuperscript{1}\textit{Joint Quantum Institute, University of Maryland, College Park, MD 20742, USA}\\\textsuperscript{2}\textit{Department of Physics, Brandeis University, Waltham, MA 02453, USA}}}

\begin{document}

\maketitle

\begin{abstract}

There is an ambiguity in how to apply the replica trick to spin glass models which have additional order parameters unrelated to spin glass order --- with respect to which quantities does one minimize vs maximize the action, and in what sequence?
Here we show that the correct procedure is to first maximize with respect to ``replica'' order parameters, and then minimize with respect to ``conventional'' order parameters.
With this result, we further elucidate the relationship between quenched free energies, annealed free energies, and replica order --- it is possible for the quenched and annealed free energies to differ even while all replica order parameters remain zero.

\end{abstract}

\tableofcontents

\section{Introduction} \label{sec:introduction}

\subsection{Opening remarks} \label{subsec:opening_remarks}

Spin Hamiltonians in which the interaction coefficients are random variables feature in various domains of theoretical physics.
They are used to model spin glasses, systems in which the magnetic moments are frozen but disordered at low temperature~\cite{Binder1986Spin,Mezard1987,Fischer1991,Mydosh1993}.
Relatedly, they have been used to establish deep connections between statistical mechanics and optimization problems in computer science~\cite{Nishimori2001,Mezard2009}.
\textit{Quantum} random Hamiltonians have also received renewed interest due to relationships to holography, quantum gravity, and non-Fermi liquids~\cite{Maldacena2016Remarks,Chowdhury2022Sachdev}.
These examples are by no means exhaustive, and countless more references can be found in those given above.

Analysis of such Hamiltonians is difficult, even for mean-field models in which every degree of freedom interacts equivalently with all others.
In this mean-field context, which is the focus of the present work, the ``replica trick'' has proven to be instrumental~\cite{Mezard1987,Castellani2005Spin,Mezard2009} (see also the recent Ref.~\cite{Charbonneau2022From} for a historical overview).
Although sometimes described as a purely mathematical (and hardly rigorous) manipulation, many physical phenomena related to ergodicity-breaking can be seen quite naturally via the replica theory.

A further aspect of random Hamiltonians that the replica trick quantifies is the distinction between ``quenched'' and ``annealed'' free energies.
For example, consider a classical Ising model (spins labeled by $i \in \{1, \cdots, N\}$) of the form
\begin{equation} \label{eq:generic_Ising_Hamiltonian}
H = \sum_{ij} J_{ij} \sigma_i \sigma_j,
\end{equation}
where $\sigma_i \in \{+1, -1\}$, and the set of couplings $\{J_{ij}\} \equiv J$ is drawn randomly from some joint probability distribution $P(J)$.
The partition function $Z(J) \equiv \textrm{Tr} e^{-\beta H}$ is clearly a function of the couplings $J$.
The quenched and annealed free energy densities, $f_{\textrm{Q}}$ and $f_{\textrm{A}}$ respectively, differ in whether one averages $\log{Z(J)}$ or $Z(J)$ itself:
\begin{equation} \label{eq:quenched_annealed_definitions}
f_{\textrm{Q}} \equiv -\lim_{N \rightarrow \infty} (N \beta)^{-1} \mathbb{E}_J \log{Z(J)}, \qquad f_{\textrm{A}} \equiv -\lim_{N \rightarrow \infty} (N \beta)^{-1} \log{\mathbb{E}_J Z(J)},
\end{equation}
where $\mathbb{E}_J \cdots \equiv \int \textrm{d}J P(J) \cdots$ denotes an average over $J$.

The quenched free energy $f_{\textrm{Q}}$ is the physically relevant quantity for disordered systems~\cite{Mezard1987,Fischer1991}, since it treats the couplings as fixed parameters when computing thermodynamic quantities and only averages over them afterwards (note in particular that derivatives of $f_{\textrm{Q}}$ yield the disorder-averaged values of observables).
Yet the annealed free energy $f_{\textrm{A}}$ can be important as well --- it is considered in the context of inference problems~\cite{Zdeborova2016Statistical} and has applications in random matrix theory~\cite{Foini2022Annealed}.
Moments of the partition function are also of interest for quantum gravity, owing to relationships between gravitational path integrals and matrix integrals, e.g., as in Ref.~\cite{saad2019jt}.
Finally, at the very least, $f_{\textrm{A}}$ serves as a simple lower bound to $f_{\textrm{Q}}$ by Jensen's inequality.

Thus it is natural to ask whether the two free energies are in fact equivalent in a given model at a given point in the phase diagram.
In the inference context, this informs the ability to perform ``quiet planting'', as discussed in Ref.~\cite{Zdeborova2016Statistical}.
It is also a practical matter in general, since the annealed free energy is more straightforward to evaluate than the quenched free energy (and see Ref.~\cite{Baldwin2020Quenched} for examples in which the two can be proven to be equal without needing to calculate the latter).
One often finds that there are distinct ``phases'' in which $f_{\textrm{Q}} = f_{\textrm{A}}$ and in which $f_{\textrm{Q}} \neq f_{\textrm{A}}$.
In particular, Ref.~\cite{Baldwin2020Quenched} gives a proof that $f_{\textrm{Q}} \neq f_{\textrm{A}}$ at low temperature in any mean-field model (whether classical or quantum) with infinite-range random interactions between \textit{local} degrees of freedom such as spins or bosons (which itself stands in contrast to the behavior in certain fermionic analogues~\cite{GurAri2018Does,Wang2019On}).

Yet in situations where $f_{\textrm{Q}} \neq f_{\textrm{A}}$, the implications for the quenched free energy itself (which we reiterate is the relevant thermodynamic quantity for disordered systems) are in fact quite subtle.
The replica trick mentioned above identifies whether $f_{\textrm{Q}} \neq f_{\textrm{A}}$ as part of a larger calculation of the quenched free energy, and it makes clear that there is a close relationship to the order parameters used to describe the (quenched) system.
The purpose of the present paper, however, is to clarify that relationship by resolving certain ambiguities in the application of the replica trick.
To the best of our knowledge, this issue has not been discussed in the literature (with one exception, which we discuss below).
Since even stating the problem requires some explanation, we feel that it is best to begin with a concrete example.

Before proceeding, we must acknowledge two caveats.
First, the results here solely concern mean-field models with Gaussian random interactions.
These models are already quite rich and of significant interest in their own right (as the above references and those therein can attest), but it would of course be valuable to consider whether and how our results extend beyond mean-field theory.
Second, as noted above, the replica trick is in general rather heuristic and certainly not rigorous --- there are situations in which it is either known or conjectured not to give correct results~\cite{Tanaka2007Moment,Mourrat2021Nonconvex}.
Our results should be viewed as similarly heuristic.
Yet given the enormous success of the replica trick in mean-field models, we do not see this as a major detraction.

\subsection{An example --- the Sherrington-Kirkpatrick model} \label{subsec:sherrington_kirkpatrick}

Consider the famous Sherrington-Kirkpatrick (SK) model for a classical Ising spin glass~\cite{Sherrington1975Solvable}:
\begin{equation} \label{eq:Sherrington_Kirkpatrick_model_definition}
H = \sum_{i < j} J_{ij} \sigma_i \sigma_j,
\end{equation}
where each $J_{ij}$ is an independent Gaussian with mean zero and variance $1/N$ (thus the interactions are infinite-range).
The annealed free energy of this model is straightforward to evaluate:
\begin{equation} \label{eq:Sherrington_Kirkpatrick_annealed_evaluation}
\mathbb{E}_J Z(J) = \sum_{\sigma} \exp{\left[ \frac{(N-1) \beta^2}{4} \right]} = \exp{\left[ \frac{(N-1) \beta^2}{4} + N \log{2} \right]},
\end{equation}
and thus
\begin{equation} \label{eq:Sherrington_Kirkpatrick_annealed_free_energy}
f_{\textrm{A}} = -\frac{1}{\beta} \log{2} - \frac{\beta}{4}.
\end{equation}

The replica trick enters for the calculation of the quenched free energy, via the mathematical identity
\begin{equation} \label{eq:replica_trick}
\mathbb{E}_J \log{Z(J)} = \lim_{n \rightarrow 0} n^{-1} \log{\mathbb{E}_J Z(J)^n}.
\end{equation}
Note that the left-hand side is precisely $-N \beta f_{\textrm{Q}}$.
\textit{For positive integer $n$}, $\mathbb{E}_J Z(J)^n$ can be evaluated without much more difficulty than $\mathbb{E}_J Z(J)$.
The result can be expressed (see Refs.~\cite{Mezard1987,Fischer1991} for details) as an integral over the off-diagonal components of matrix $Q_{\alpha \alpha'}$, where $\alpha, \alpha' \in \{1, \cdots, n\}$ label the different factors (``replicas'') of $Z(J)$:
\begin{equation} \label{eq:Sherrington_Kirkpatrick_partition_integral}
\mathbb{E}_J Z(J)^n \sim \int \textrm{d}Q \exp{\big[ -Nn \beta S_n(Q) \big]},
\end{equation}
where $\int \textrm{d}Q$ is shorthand (up to unimportant prefactors) for the integral over all components of $Q$, with effective replicated action
\begin{equation} \label{eq:Sherrington_Kirkpatrick_replicated_action}
S_n(Q) \equiv -\frac{\beta}{4} + \frac{\beta}{2n} \sum_{\alpha < \alpha'} Q_{\alpha \alpha'}^2 - \frac{1}{n \beta} \log{\textrm{Tr} \exp{\left[ \beta^2 \sum_{\alpha < \alpha'} Q_{\alpha \alpha'} \sigma^{\alpha} \sigma^{\alpha'} \right]}}.
\end{equation}
Eq.~\eqref{eq:Sherrington_Kirkpatrick_partition_integral} is then evaluated by saddle point at large $N$, and the saddle-point value of $S_n(Q)$ in the limit $n \rightarrow 0$ is precisely $f_{\textrm{Q}}$ (see Eq.~\eqref{eq:replica_trick}).
Note that the equations determining the saddle points can be written
\begin{equation} \label{eq:Sherrington_Kirkpatrick_saddle_point_equations}
Q_{\alpha \alpha'} = \big< \sigma^{\alpha} \sigma^{\alpha'} \big>_{\textrm{eff}},
\end{equation}
where $\langle \, \cdot \, \rangle_{\textrm{eff}}$ denotes a thermal expectation value at unit temperature with respect to the single-site but many-replica Hamiltonian $H_{\textrm{eff}} \equiv -\beta^2 \sum_{\alpha < \alpha'} Q_{\alpha \alpha'} \sigma^{\alpha} \sigma^{\alpha'}$.
We can thus interpret $Q_{\alpha \alpha'}$ as the order parameter characterizing the degree of correlation between replicas $\alpha$ and $\alpha'$.

First note that the annealed free energy is recovered simply by setting $Q = 0$ in Eq.~\eqref{eq:Sherrington_Kirkpatrick_replicated_action}.
This makes sense --- averaging over $J$ introduces terms into the action which couple the replicas and (due to the mean-field nature of the model) can be expressed solely in terms of $Q$.
Those terms vanish when $Q = 0$, meaning that $\mathbb{E}_J Z(J)^n \sim [\mathbb{E}_J Z(J)]^n$ and the right-hand side of Eq.~\eqref{eq:replica_trick} becomes $\log{\mathbb{E}_J Z(J)}$.

Determining the quenched free energy requires considering all values of $Q$, however.
One of the more mysterious aspects of the replica trick is that Eq.~\eqref{eq:Sherrington_Kirkpatrick_partition_integral} is dominated as $n \rightarrow 0$ by the saddle point which \textit{maximizes} $S_n(Q)$, i.e., the saddle point which seems to give the \textit{smallest} contribution to $\mathbb{E}_J Z(J)^n$.
As bizarre as it is, the final expression for the quenched free energy has been rigorously verified through independent (but far more technical and opaque) means~\cite{Bovier2006,Talagrand2011a,Talagrand2011b,Panchenko2013}, so one can safely accept this prescription of maximizing the action.
We can thus summarize the expressions for the two free energies as (with $n \rightarrow 0$ implied)
\begin{equation} \label{eq:simple_quenched_annealed_expressions}
f_{\textrm{Q}} = \max_Q S_n(Q), \qquad f_{\textrm{A}} = S_n(0).
\end{equation}
Note that the two are equal if and only if $S_n(Q)$ is maximized at $Q = 0$.

When the maximum is at $Q \neq 0$, we refer to the system as having ``replica order''.
An important subtlety is that replica order does \textit{not} necessarily imply spin glass order --- it is standard to identify spin glass order with saddle points that further break the permutation symmetry between the $n$ replicas, corresponding physically to broken ergodicity\footnote{We refer to standard textbooks~\cite{Mezard1987,Castellani2005Spin,Mezard2009} for a full discussion of this point, but briefly, broken ergodicity and the existence of multiple equilibrium states implies that some replicas may lie in the same state while others lie in different states, hence a lack of permutation symmetry among replicas.}.
That said, spin glass order is a specific type of replica order and is often found in practice (such as in the SK model at low temperature).

There is evidently a relationship between whether $f_{\textrm{Q}} = f_{\textrm{A}}$ and replica order.
\textit{In the SK model}, this relationship is quite straightforward --- $f_{\textrm{Q}} = f_{\textrm{A}}$ if and only if there is no replica order.
One can see why from Eq.~\eqref{eq:Sherrington_Kirkpatrick_saddle_point_equations} --- $Q = 0$ signifies that the replicas are uncorrelated, and thus $\mathbb{E}_J Z(J)^n$ factors into $[\mathbb{E}_J Z(J)]^n$.
Given this, it is tempting to extrapolate and assume that $Q = 0$ always implies $f_{\textrm{Q}} = f_{\textrm{A}}$.
However, \textit{one of our main results is that this is not true --- it is possible for the two free energies to differ even in the absence of replica order}.
To see why, we must turn to more complex models in this work.

\subsection{Summary of results} \label{subsec:summary_results}

The SK and related models are special in that they \textit{only} have spin glass order parameters.
More complicated models (even still infinite-range) may have additional order parameters unrelated to spin glass order.
The replicated partition function will then take the form (compare to Eq.~\eqref{eq:Sherrington_Kirkpatrick_partition_integral})
\begin{equation} \label{eq:generic_replicated_path_integral}
\mathbb{E}_J Z(J)^n = \int \textrm{d}R \textrm{d}Q \exp{\big[ -Nn \beta S_n(R, Q) \big]},
\end{equation}
where $Q$ denotes the set of order parameters characterizing inter-replica correlations as in the SK model, and $R$ denotes those characterizing single-replica properties.
More precisely, the variables $Q_{\alpha \alpha'}$ will carry two replica indices\footnote{In fact, the quantities $Q$ will carry only two replica indices (as opposed to higher numbers as well) only when the couplings $J_{ij}$ are Gaussian-distributed and enter linearly into the Hamiltonian, such as in Eq.~\eqref{eq:Sherrington_Kirkpatrick_model_definition}. This is by far the situation most considered in the literature, at least for infinite-range models, and we focus on it as well.} and the variables $R_{\alpha}$ will carry a single replica index.
We refer to the two-index quantities as ``replica'' order parameters and the one-index quantities as ``conventional'' order parameters.
The action $S_n(R, Q)$ of course depends on the specific model under consideration.
We give explicit examples in the following sections (and note that this form holds only in mean-field models).

While in principle one again simply has to identify the dominant saddle point, the presence of conventional order parameters makes this task even more delicate.
We know that the replica trick entails \textit{maximizing} the action with respect to $Q$, but since the conventional order parameters are unrelated to inter-replica correlations, one would expect to still \textit{minimize} with respect to $R$.
Yet the operations of maximizing over $Q$ and minimizing over $R$ do not generically commute, and so one still has to determine in which order to perform the two (if this really is the correct procedure to follow).
This is the ambiguity in the replica trick that we alluded to above.

In this paper, we show that the correct procedure is to first maximize with respect to $Q$, giving an effective action solely in terms of $R$, and then minimize with respect to $R$.
In other words,
\begin{equation} \label{eq:min_max_prescription}
-\lim_{N \rightarrow \infty} (N \beta)^{-1} \mathbb{E}_J \log{Z(J)} = \min_R \Big[ \max_Q \big[ S_n(R, Q) \big] \Big],
\end{equation}
which we refer to as the ``min-max'' prescription.
To the best of our knowledge, this prescription has not yet been articulated in the literature.
While one can often succeed in selecting the correct saddle point on physical grounds~\cite{Goldschmidt1990Solvable,Nieuwenhuizen1998Quantum,Cugliandolo2000From,Cugliandolo2001Imaginary}, it is nonetheless desirable to have an explicit procedure such as Eq.~\eqref{eq:min_max_prescription} which does not require independent insight.
This is especially true given the renewed interest in quantum models, which generically contain conventional order parameters almost by definition\footnote{When expressed as path integrals, the degrees of freedom in quantum models have imaginary-time dependence, meaning that the (single-replica) imaginary-time correlation function appears as a conventional order parameter once averaging over disorder.}.

The min-max prescription explains how the quenched and annealed free energies can differ even without replica order.
Since the maximization over $Q$ occurs separately for each value of $R$, the maximum is at a function $Q_c(R)$, and we can write $f_{\textrm{Q}} = \min_R S_n(R, Q_c(R))$.
On the other hand, the annealed free energy is still recovered by setting $Q = 0$, but now for all values of $R$ and with the minimization over $R$ remaining.
We thus have that
\begin{equation} \label{eq:complex_quenched_annealed_expressions}
f_{\textrm{Q}} = \min_R S_n \big( R, Q_c(R) \big), \qquad f_{\textrm{A}} = \min_R S_n \big( R, 0 \big).
\end{equation}
Denote the values of $R$ at which the two minima are obtained by $R_{\textrm{Q}}$ and $R_{\textrm{A}}$ respectively.
Clearly we have that $f_{\textrm{Q}} = f_{\textrm{A}}$ if $Q_c(R) = 0$ for all $R$, and slightly more generally\footnote{To see this, note for all $R$ and $Q$, we have by definition that $S_n(R, Q_c(R)) \geq S_n(R, 0) \geq S_n(R_{\textrm{A}}, 0)$. Thus if $Q_c(R_{\textrm{A}}) = 0$, the minimum of $S_n(R, Q_c(R))$ must be at $R_{\textrm{A}}$, meaning $f_{\textrm{Q}} = S_n(R_{\textrm{A}}, 0) = f_{\textrm{A}}$.}, $f_{\textrm{Q}} = f_{\textrm{A}}$ if $Q_c(R_{\textrm{A}}) = 0$.
Yet replica order refers to whether $Q_c(R_{\textrm{Q}}) = 0$.
It can very well be that $R_{\textrm{Q}} \neq R_{\textrm{A}}$, and there is nothing preventing one from having $Q_c(R_{\textrm{Q}}) = 0$ even though $Q_c(R_{\textrm{A}}) \neq 0$.
In such a situation, the two free energies differ (since $R_{\textrm{Q}} \neq R_{\textrm{A}}$) but there is no replica order (since $Q_c(R_{\textrm{Q}}) = 0$).

One would be hard-pressed to justify having both $f_{\textrm{Q}} \neq f_{\textrm{A}}$ and $Q = 0$ without the min-max prescription in mind, for it is still true that the replicated action $S_n(R, Q)$ reduces to that of the annealed calculation by setting $Q = 0$.
One then has to explain how the same action can yield different results in the quenched and annealed situations.
For example, if one were to perform the optimizations in the opposite order --- $\max_Q \min_R S_n(R, Q)$ --- then it would be the case that $Q = 0$ implies $f_{\textrm{Q}} = f_{\textrm{A}}$.
The min-max prescription avoids this by having a separate maximization over $Q$ for each value of $R$, so that one can have $Q = 0$ for some but not all $R$.

It remains to justify the min-max prescription, Eq.~\eqref{eq:min_max_prescription}.
We do so in Secs.~\ref{sec:random_energy_models} and~\ref{sec:Aizenman_Sims_Starr}.
Sec.~\ref{sec:random_energy_models} studies systems consisting of a generic collection of random energy models (REMs).
These models are useful because they can be analyzed both through replicas and by direct calculation.
By considering a sufficiently broad class of REMs, we show that only the min-max prescription gives the correct quenched free energy in all cases.
This same approach was used in Ref.~\cite{Mottishaw1986First} --- the only work of which we are aware to explicitly consider how to apply the replica trick to models with conventional order parameters --- albeit applied to a much more restricted class of models (and we in fact disagree with certain of their conclusions, as discussed below).
Sec.~\ref{sec:Aizenman_Sims_Starr} then considers the rigorous theory of infinite-range spin glasses, generalized to models with conventional order parameters.
Although a complete analysis is beyond the scope of this work, we show that the ``Aizenman-Sims-Starr'' scheme, which is a starting point for many rigorous results, can be adapted to include conventional order parameters precisely by following the min-max prescription.
Examples of this can already be found in the mathematical literature~\cite{Panchenko2018FreeA,Panchenko2018FreeB,Camilli2022Inference}, and the purpose of this section is to establish the relevance of those results in the present context and generalize them further.

Beforehand, we revisit the replica theory of the transverse-field $p$-spin model in Sec.~\ref{sec:p_spin}.
This is important for two reasons.
First, for all that we have already said, we have not yet demonstrated that different prescriptions in the replica theory can yield different results for a model of independent interest.
The transverse-field $p$-spin model turns out to be such a system.
It has been well-studied both to understand the effects of quantum fluctuations on spin glass phases~\cite{Goldschmidt1990Solvable,Dobrosavljevic1990Expansion,Nieuwenhuizen1998Quantum,Baldwin2017Clustering,Biroli2021OutOfEquilibrium} and in the context of quantum computing~\cite{Jorg2008Simple,Baldwin2018Quantum,Smelyanskiy2020Nonergodic}.

Second, the transverse-field $p$-spin model also provides an example in which $f_{\textrm{Q}} \neq f_{\textrm{A}}$ without there being any replica order.
The proof in Ref.~\cite{Baldwin2020Quenched} applies to this model (with a trivial generalization\footnote{Strictly speaking, Ref.~\cite{Baldwin2020Quenched} considers models that consist only of infinite-range Gaussian interactions, but it is clear from the proof technique that the result holds regardless of any other terms in the Hamiltonian as well (as long as those terms are independent of the random interactions). In short, averaging the partition function over the Gaussian couplings always gives a factor with exponent going as $\beta^2$ (see Eq.~\eqref{eq:Sherrington_Kirkpatrick_annealed_evaluation} for an example), irrespective of what other terms are in the Hamiltonian. This translates to an upper bound on the annealed free energy going as $-\beta$, hence $f_{\textrm{A}} \rightarrow -\infty$ as $\beta \rightarrow \infty$, again regardless of any additional terms. Yet in systems with a finite local Hilbert space, such behavior cannot occur in the quenched free energy. See Ref.~\cite{Baldwin2020Quenched} for further details.}), meaning that $f_{\textrm{Q}} \neq f_{\textrm{A}}$ at low temperature, \textit{regardless of the transverse field strength}.
On the other hand, the commonly-accepted phase diagram of this model is completely featureless, and in particular has $Q = 0$ at all temperatures, for fields exceeding a critical value $\Gamma_c$~\cite{Goldschmidt1990Solvable,Dobrosavljevic1990Expansion,Nieuwenhuizen1998Quantum}.
This pair of observations was in fact the original motivation for the present work.
We have argued above that the min-max prescription is precisely what allows for both to hold simultaneously, and it is satisfying to see this explicitly in the transverse-field $p$-spin model at large field.

\section{Transverse-field p-spin model} \label{sec:p_spin}

The transverse-field $p$-spin model consists of all-to-all random interactions between the $z$-components of $N$ spin-1/2s (plus a uniform transverse field):
\begin{equation} \label{eq:TFpS_definition}
H_p = \sum_{(i_1 \cdots i_p)} J_{i_1 \cdots i_p} \hat{\sigma}_{i_1}^z \cdots \hat{\sigma}_{i_p}^z - \Gamma \sum_i \hat{\sigma}_i^x,
\end{equation}
where the first sum is over all tuples of $p$ indices (i.e., $i_1 < \cdots < i_p$), and each $J_{i_1 \cdots i_p}$ is Gaussian-distributed with mean 0 and variance $p!/2N^{p-1}$.
Here $\hat{\sigma}^a$ denotes the $a$-component of the Pauli spin operator, while $\sigma$ without a hat will denote a classical variable taking values $\pm 1$.
The integer $p$ is considered another parameter of the model.
We shall specifically take the large-$p$ limit, since this allows for explicit final expressions and justifies the ansatz used at a later stage, but the same qualitative features can be confirmed at finite $p$ (at least within that ansatz) using Eq.~\eqref{eq:p_spin_simplified_effective_action} below.
We follow Ref.~\cite{Goldschmidt1990Solvable} in our derivation of the replicated effective action, then show that different prescriptions yield different phase diagrams.

To evaluate the $n$'th moment of the partition function, we first express it as a classical partition function following the usual Suzuki-Trotter procedure:
\begin{equation} \label{eq:p_spin_partition_function}
\mathbb{E}_J \big( \textrm{Tr} e^{-\beta H_p} \big)^n = \mathbb{E}_J \textrm{Tr} \exp{\left[ -\frac{\beta}{M} \sum_{\tau = 1}^M \sum_{\alpha = 1}^n \sum_{(i_1 \cdots i_p)} J_{i_1 \cdots i_p} \sigma_{i_1}^{\alpha}(\tau) \cdots \sigma_{i_p}^{\alpha}(\tau) + \sum_{\alpha = 1}^n \sum_{i = 1}^N H_{\Gamma}(\sigma_i^{\alpha}) \right]},
\end{equation}
where $\sigma_i^{\alpha}(\tau)$ denotes the value of spin $i$ on replica $\alpha$ at imaginary-time slice $\tau \in \{1, \cdots, M\}$, and
\begin{equation} \label{eq:Trotter_term}
H_{\Gamma}(\sigma_i^{\alpha}) \equiv \sum_{\tau = 1}^M \left( \frac{\sigma_i^{\alpha}(\tau) \sigma_i^{\alpha}(\tau + 1)}{2} \log{\coth{\frac{\beta \Gamma}{M}}} + \frac{1}{2} \log{\frac{1}{2} \sinh{\frac{2 \beta \Gamma}{M}}} \right).
\end{equation}
Carrying out the average over disorder gives
\begin{equation} \label{eq:p_spin_partition_function_evaluation_v1}
\begin{aligned}
\mathbb{E}_J \big( \textrm{Tr} e^{-\beta H_p} \big)^n &= \textrm{Tr} \exp{\left[ \sum_{(i_1 \cdots i_p)} \frac{p! \beta^2}{4N^{p-1} M^2} \sum_{\alpha \alpha'} \sum_{\tau \tau'} \sigma_{i_1}^{\alpha}(\tau) \sigma_{i_1}^{\alpha'}(\tau') \cdots \sigma_{i_p}^{\alpha}(\tau) \sigma_{i_p}^{\alpha'}(\tau') + \sum_{\alpha i} H_{\Gamma}(\sigma_i^{\alpha}) \right]} \\
&\sim \textrm{Tr} \exp{\left[ \frac{N \beta^2}{4M^2} \sum_{\alpha \alpha'} \sum_{\tau \tau'} \left( \frac{1}{N} \sum_i \sigma_i^{\alpha}(\tau) \sigma_i^{\alpha'}(\tau') \right)^p + \sum_{\alpha i} H_{\Gamma}(\sigma_i^{\alpha}) \right]},
\end{aligned}
\end{equation}
introducing order parameters gives
\begin{equation} \label{eq:p_spin_partition_function_evaluation_v2}
\begin{aligned}
\mathbb{E}_J \big( \textrm{Tr} e^{-\beta H_p} \big)^n &\sim \int \mathcal{D}R \mathcal{D}Q \exp{\left[ \frac{N \beta^2}{4M^2} \sum_{\alpha} \sum_{\tau \tau'} R_{\alpha}(\tau, \tau')^p + \frac{N \beta^2}{4M^2} \sum_{\alpha \neq \alpha'} \sum_{\tau \tau'} Q_{\alpha \alpha'}(\tau, \tau')^p \right]} \\
& \qquad \cdot \textrm{Tr} \exp{\left[ \sum_{\alpha i} H_{\Gamma}(\sigma_i^{\alpha}) \right]} \prod_{\alpha} \prod_{\tau < \tau'} \delta \left( R_{\alpha}(\tau, \tau')  - \frac{1}{N} \sum_i \sigma_i^{\alpha}(\tau) \sigma_i^{\alpha}(\tau') \right) \\
& \qquad \qquad \cdot \prod_{\alpha < \alpha'} \prod_{\tau \tau'} \delta \left( Q_{\alpha \alpha'}(\tau, \tau') - \frac{1}{N} \sum_i \sigma_i^{\alpha}(\tau) \sigma_i^{\alpha'}(\tau') \right),
\end{aligned}
\end{equation}
and introducing Lagrange multipliers gives
\begin{equation} \label{eq:p_spin_partition_function_evaluation_v3}
\begin{aligned}
\mathbb{E}_J \big( \textrm{Tr} e^{-\beta H_p} \big)^n &\sim \int \mathcal{D}R \mathcal{D}Q \mathcal{D}K \mathcal{D}\Lambda \exp{\left[ \frac{N \beta^2}{4M^2} \sum_{\alpha} \sum_{\tau \tau'} R_{\alpha}(\tau, \tau')^p + \frac{N \beta^2}{4M^2} \sum_{\alpha \neq \alpha'} \sum_{\tau \tau'} Q_{\alpha \alpha'}(\tau, \tau')^p \right]} \\
& \quad \cdot \exp{\left[ -\frac{N \beta^2}{2M^2} \sum_{\alpha} \sum_{\tau < \tau'} K_{\alpha}(\tau, \tau') R_{\alpha}(\tau, \tau') - \frac{N \beta^2}{2M^2} \sum_{\alpha < \alpha'} \sum_{\tau \tau'} \Lambda_{\alpha \alpha'}(\tau, \tau') Q_{\alpha \alpha'}(\tau, \tau') \right]} \\
& \quad \quad \cdot \left( \textrm{Tr} \exp{\left[ \sum_{\alpha} H_{\Gamma}(\sigma^{\alpha}) + \frac{\beta^2}{2M^2} \sum_{\alpha} \sum_{\tau < \tau'} K_{\alpha}(\tau, \tau') \sigma^{\alpha}(\tau) \sigma^{\alpha}(\tau') \right]} \right. \\
& \qquad \qquad \qquad \qquad \qquad \cdot \left. \exp{\left[ \frac{\beta^2}{2M^2} \sum_{\alpha < \alpha'} \sum_{\tau \tau'} \Lambda_{\alpha \alpha'}(\tau, \tau') \sigma^{\alpha}(\tau) \sigma^{\alpha'}(\tau') \right]} \right)^N,
\end{aligned}
\end{equation}
where $\int \mathcal{D}R \mathcal{D}Q \mathcal{D}K \mathcal{D}\Lambda$ denotes the integral over all components of $R$, $Q$, $K$, $\Lambda$.
See Ref.~\cite{Goldschmidt1990Solvable}, as well as Refs.~\cite{Gardner1985Spin,Castellani2005Spin,Mezard2009}, for details of each step above.

The remaining integrals can be evaluated by saddle point at large $N$.
At the dominant saddle point, $R_{\alpha}(\tau, \tau')$ can be interpreted\footnote{Note that Eqs.~\eqref{eq:p_spin_R_interpretation} and~\eqref{eq:p_spin_Q_interpretation} follow directly from Eq.~\eqref{eq:p_spin_partition_function_evaluation_v2}, without needing to introduce the effective single-spin Hamiltonian seen in Eq.~\eqref{eq:p_spin_partition_function_evaluation_v3}. The saddle-point equations give $R$ and $Q$ \textit{additional} interpretations as expectation values of $\sigma^{\alpha}(\tau) \sigma^{\alpha}(\tau')$ and $\sigma^{\alpha}(\tau) \sigma^{\alpha'}(\tau')$ with respect to that single-spin Hamiltonian, but we will not need those interpretations for the present analysis.} as the (imaginary-time) autocorrelation function of replica $\alpha$:
\begin{equation} \label{eq:p_spin_R_interpretation}
R_{\alpha}(\tau, \tau') = \frac{1}{N} \sum_i \big< \hat{\sigma}_i^{\alpha z}(\tau) \hat{\sigma}_i^{\alpha z}(\tau') \big>,
\end{equation}
where $\langle \, \cdot \, \rangle$ denotes the thermal expectation value with respect to the original Hamiltonian $H_p$, understood to act independently on each replica, and $\hat{\sigma}_i^{\alpha z}(\tau) \equiv e^{\tau H_p} \hat{\sigma}_i^{\alpha z} e^{-\tau H_p}$ (with time-ordering implied).
Similarly, $Q_{\alpha \alpha'}(\tau, \tau')$ can be interpreted as the inter-replica correlation function:
\begin{equation} \label{eq:p_spin_Q_interpretation}
Q_{\alpha \alpha'}(\tau, \tau') = \frac{1}{N} \sum_i \big< \hat{\sigma}_i^{\alpha z}(\tau) \big> \big< \hat{\sigma}_i^{\alpha' z}(\tau') \big>.
\end{equation}
The quantities $K$ and $\Lambda$ are the Lagrange multipliers corresponding to $R$ and $Q$ respectively (see the discussion in App.~\ref{app:conventional_extremizing}).

Note that two of the saddle-point equations are simply
\begin{equation} \label{eq:p_spin_easy_saddle_points}
K_{\alpha}(\tau, \tau') = pR_{\alpha}(\tau, \tau')^{p-1}, \qquad \Lambda_{\alpha \alpha'}(\tau, \tau') = pQ_{\alpha \alpha'}(\tau, \tau')^{p-1},
\end{equation}
which allows us to eliminate $K$ and $\Lambda$ (although see App.~\ref{app:conventional_extremizing}).
Furthermore, since all replicas are equivalent and each has a separate time-translation invariance, we can take as an ansatz solutions of the form
\begin{equation} \label{eq:p_spin_saddle_point_structure}
R_{\alpha}(\tau, \tau') = R(\tau - \tau'), \qquad Q_{\alpha \alpha'}(\tau, \tau') = Q_{\alpha \alpha'}.
\end{equation}
This is especially justified given the expressions in Eqs.~\eqref{eq:p_spin_R_interpretation} and~\eqref{eq:p_spin_Q_interpretation}.
Thus the partition function reduces to the form in Eq.~\eqref{eq:generic_replicated_path_integral}:
\begin{equation} \label{eq:p_spin_effective_partition_function}
\mathbb{E}_J \big( \textrm{Tr} e^{-\beta H_p} \big)^n \sim \int \mathcal{D}R \mathcal{D}Q \exp{\big[ -Nn \beta S_n(R, Q) \big]},
\end{equation}
with effective action
\begin{equation} \label{eq:p_spin_effective_action}
\begin{aligned}
S_n(R, Q) &= \frac{(p-1) \beta}{4M} \sum_{\tau} R(\tau)^p + \frac{(p-1) \beta}{4n} \sum_{\alpha \neq \alpha'} Q_{\alpha \alpha'}^p \\
& \qquad - \frac{1}{n \beta} \log{\textrm{Tr} \exp{\left[ \sum_{\alpha} H_{\Gamma}(\sigma^{\alpha}) + \frac{p \beta^2}{4M^2} \sum_{\alpha} \sum_{\tau \tau'} R(\tau - \tau')^{p-1} \sigma^{\alpha}(\tau) \sigma^{\alpha}(\tau') \right]}} \\
& \qquad \qquad \qquad \qquad \cdot \exp{\left[ \frac{p \beta^2}{4M^2} \sum_{\alpha \neq \alpha'} \sum_{\tau \tau'} Q_{\alpha \alpha'}^{p-1} \sigma^{\alpha}(\tau) \sigma^{\alpha'}(\tau') \right]}.
\end{aligned}
\end{equation}

Extremizing Eq.~\eqref{eq:p_spin_effective_action} with respect to $R$ and $Q$ is still not quite tractable (even without the subtleties of taking $n \rightarrow 0$).
Thus we consider the large-$p$ limit, in which a static 1RSB ansatz is known to be valid~\cite{Goldschmidt1990Solvable,Nieuwenhuizen1998Quantum}: take $R(\tau - \tau') = R$ independent of time, and divide the replicas into $n/m$ groups of $m$ replicas each, such that $Q_{\alpha \alpha'} = Q$ for $\alpha$ and $\alpha'$ in the same group and $Q_{\alpha \alpha'} = 0$ otherwise.
In the $n \rightarrow 0$ limit, $m$ is taken to lie between 0 and 1, with $m < 1$ and $Q > 0$ corresponding to spin glass order~\cite{Mezard1987,Fischer1991,Mezard2009}.
Following some Hubbard-Stratonovich transformations and further algebra --- see Ref.~\cite{Goldschmidt1990Solvable} --- the effective action becomes
\begin{equation} \label{eq:p_spin_simplified_effective_action}
\begin{aligned}
S_n(R, m, Q) &= \frac{(p-1) \beta}{4} \Big( R^p + (m-1) Q^p \Big) \\
& \qquad - \frac{1}{m \beta} \log{\int \frac{\textrm{d}y}{\sqrt{\pi}} e^{-y^2} \left( \int \frac{\textrm{d}z}{\sqrt{\pi}} e^{-z^2} 2 \cosh{\beta \sqrt{h(y, z)^2 + \Gamma^2}} \right)^m},
\end{aligned}
\end{equation}
where $h(y, z) \equiv y \sqrt{pQ^{p-1}} + z \sqrt{p(R^{p-1} - Q^{p-1})}$.
The large-$p$ limit can now be taken depending on the values of $R$ and $Q$:
\begin{itemize}
    \item $pR^{p-1} \rightarrow 0$, $pQ^{p-1} \rightarrow 0$: The action can be expanded as
    \begin{equation} \label{eq:p_spin_QPM_action_expansion}
    S_n(R, m, Q) \sim -\frac{1}{\beta} \log{2 \cosh{\beta \Gamma}} + \frac{(p-1) \beta}{4} R^p - \frac{p}{4 \Gamma} R^{p-1} \tanh{\beta \Gamma} + \frac{(p-1) \beta}{4} (m-1) Q^p.
    \end{equation}
    To leading order, it is simply $S_n \sim -\beta^{-1} \log{2 \cosh{\beta \Gamma}}$.
    There is a saddle point at
    \begin{equation} \label{eq:p_spin_QPM_saddle_point}
    R \sim \frac{1}{\beta \Gamma} \tanh{\beta \Gamma}, \qquad m = 1, \qquad Q = 0,
    \end{equation}
    which does not have spin glass order (technically all values of $Q$ are degenerate at $m = 1$, but the fact that $m = 1$ means that there is no spin glass order regardless).
    \item $pR^{p-1} \rightarrow \infty$, $pQ^{p-1} \rightarrow 0$: The action can be expanded as
    \begin{equation} \label{eq:p_spin_CPM_action_expansion}
    S_n(R, m, Q) \sim -\frac{1}{\beta} \log{2} + \frac{(p-1) \beta}{4} R^p - \frac{p \beta}{4} R^{p-1} - \frac{\Gamma^2}{p \beta R^{p-1}} + \frac{(p-1) \beta}{4} (m-1) Q^p.
    \end{equation}
    This regime also yields a saddle point at
    \begin{equation} \label{eq:p_spin_CPM_saddle_point}
    R \sim 1 - \frac{4 \Gamma^2}{p^2 \beta^2}, \qquad m = 1, \qquad Q = 0,
    \end{equation}
    near which $S_n \sim -\beta^{-1} \log{2} - \beta/4$ to leading order.
    \item $pR^{p-1} \rightarrow \infty$, $pQ^{p-1} \rightarrow \infty$ (with $pR^{p-1} - pQ^{p-1} \not \rightarrow \infty$): The action can be expanded as
    \begin{equation} \label{eq:p_spin_SG_action_expansion}
    \begin{aligned}
    S_n(R, m, Q) &\sim -\frac{1}{m \beta} \log{2} + \frac{(p-1) \beta}{4} R^p - \frac{p \beta}{4} R^{p-1} + \frac{(p-1) \beta}{4} (m-1) Q^p - \frac{p \beta}{4} (m-1) Q^{p-1} \\
    &\qquad \qquad \qquad \qquad - \frac{\Gamma^2}{mp \beta Q^{p-1}} + \frac{\Gamma^2}{m^2 p \beta Q^{2p-2}} \big( R^{p-1} - Q^{p-1} \big).
    \end{aligned}
    \end{equation}
    If $\beta < 2 \sqrt{\log{2}}$, then there is no saddle point with respect to $m$, meaning the maximum is at $m = 1$ and this case reduces to the previous one.
    Yet if $\beta > 2 \sqrt{\log{2}}$, then there is an additional saddle point at
    \begin{equation} \label{eq:p_spin_SG_saddle_point}
    R \sim 1 - \frac{2 \Gamma^2}{p^2 \beta \sqrt{\log{2}}}, \qquad m \sim \frac{2 \sqrt{\log{2}}}{\beta}, \qquad Q \sim 1 - \frac{\Gamma^2}{p^2 \log{2}},
    \end{equation}
    with action $S_n \sim -\sqrt{\log{2}}$.
\end{itemize}
We have thus identified three distinct saddle points.
Eq.~\eqref{eq:p_spin_QPM_saddle_point} is termed the ``quantum paramagnetic'' (QPM) saddle point, Eq.~\eqref{eq:p_spin_CPM_saddle_point} is termed the ``classical paramagnetic'' (CPM) saddle point, and Eq.~\eqref{eq:p_spin_SG_saddle_point} (when it exists) is termed the ``spin glass'' (SG) saddle point.
Note that the QPM and CPM saddle points do not have spin glass order and differ only in the value of $R$, whereas CPM and SG have similar values of $R$ but differ in the presence of spin glass order.
The corresponding QPM, CPM, and SG phases are those portions of the phase diagram in which each saddle point is dominant.

\begin{figure}[t]
\centering
\includegraphics[width=0.7\textwidth]{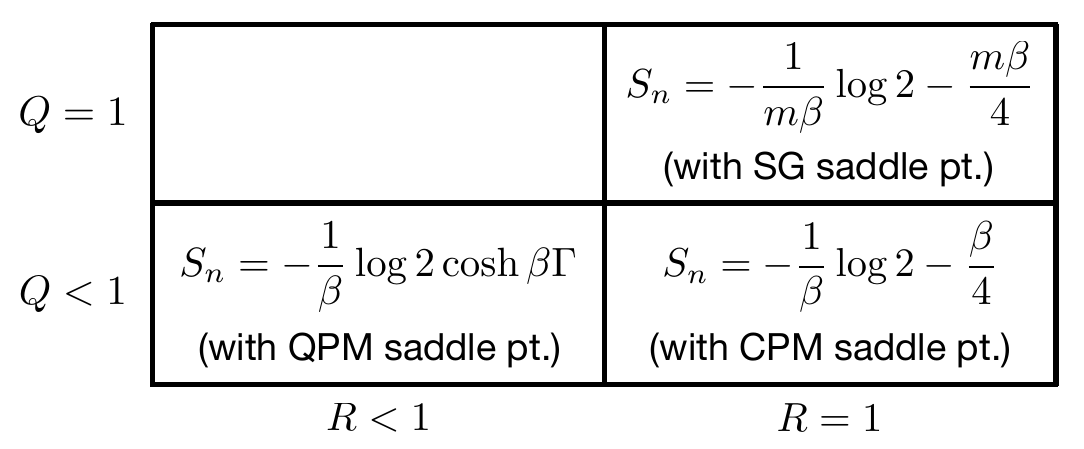}
\caption{Effective action of the transverse-field $p$-spin model to leading order at large $p$. Rows and columns indicate the values of $R$ and $Q$ (the action is independent of $R$ and/or $Q$ to leading order whenever each is strictly less than 1), and any $m$-dependence is indicated in the table entry. Subleading terms, given in the main text but not here, yield distinct saddle points in each of the three regions (the SG saddle point exists only for $\beta > 2 \sqrt{\log{2}}$). QPM stands for the ``quantum paramagnetic'' saddle point, CPM for ``classical paramagnetic'', and SG for ``spin glass''.}
\label{fig:REM_action}
\end{figure}

The values of the action in these three regions are illustrated in Fig.~\ref{fig:REM_action} (note that we do not consider $R < Q$ on physical grounds --- the correlation between replicas should not exceed the imaginary-time correlation within a single replica).
The question remains of how to extremize the action.
Here we consider six possibilities\footnote{This is not an exhaustive list. Barring a rigorous derivation (which this paper does not provide), there is always the possibility that a more complicated prescription may be correct and simply happen to reduce to the min-max prescription in the cases considered here. We cannot rule this out, and encourage further investigation.}:
\begin{itemize}
    \item Min-max prescription:
    \begin{equation} \label{eq:min_max_definition}
    f = \min_R \max_{m,Q} S_n(R, m, Q).
    \end{equation}
    \item Max-min prescription:
    \begin{equation} \label{eq:max_min_definition}
    f = \max_{m,Q} \min_R S_n(R, m, Q).
    \end{equation}
    \item Full minimization:
    \begin{equation} \label{eq:full_min_definition}
    f = \min_{R,m,Q} S_n(R, m, Q).
    \end{equation}
    \item Full maximization:
    \begin{equation} \label{eq:full_max_definition}
    f = \max_{R,m,Q} S_n(R, m, Q).
    \end{equation}
    \item Local stability (minimization): Among the two (for $\beta < 2 \sqrt{\log{2}}$) or three (for $\beta > 2 \sqrt{\log{2}}$) stable saddle points identified above, select the one with the lowest free energy.
    \item Local stability (maximization): Among those same saddle points, select the one with the highest free energy.
\end{itemize}
As stated in Sec.~\ref{sec:introduction}, we argue for Eq.~\eqref{eq:min_max_definition}, the min-max prescription.
Eq.~\eqref{eq:max_min_definition} shares the sensible behavior of minimizing over conventional and maximizing over replica order parameters, and differs only in the order in which those two are performed.
Eqs.~\eqref{eq:full_min_definition} and~\eqref{eq:full_max_definition} strike us as less justifiable, and we include them mainly for comparison.
Minimizing among the stable saddle points has been advocated for previously~\cite{Mottishaw1986First} (we discuss this prescription further in Sec.~\ref{sec:random_energy_models}), and we include the corresponding maximization for completeness as well.
Note that these last two are different than fully minimizing or maximizing over all order parameters, since the stable saddle points are local \textit{minima} with respect to $R$ but \textit{maxima} with respect to $m$ and $Q$.

\begin{figure}[t]
\centering
\includegraphics[width=0.8\textwidth]{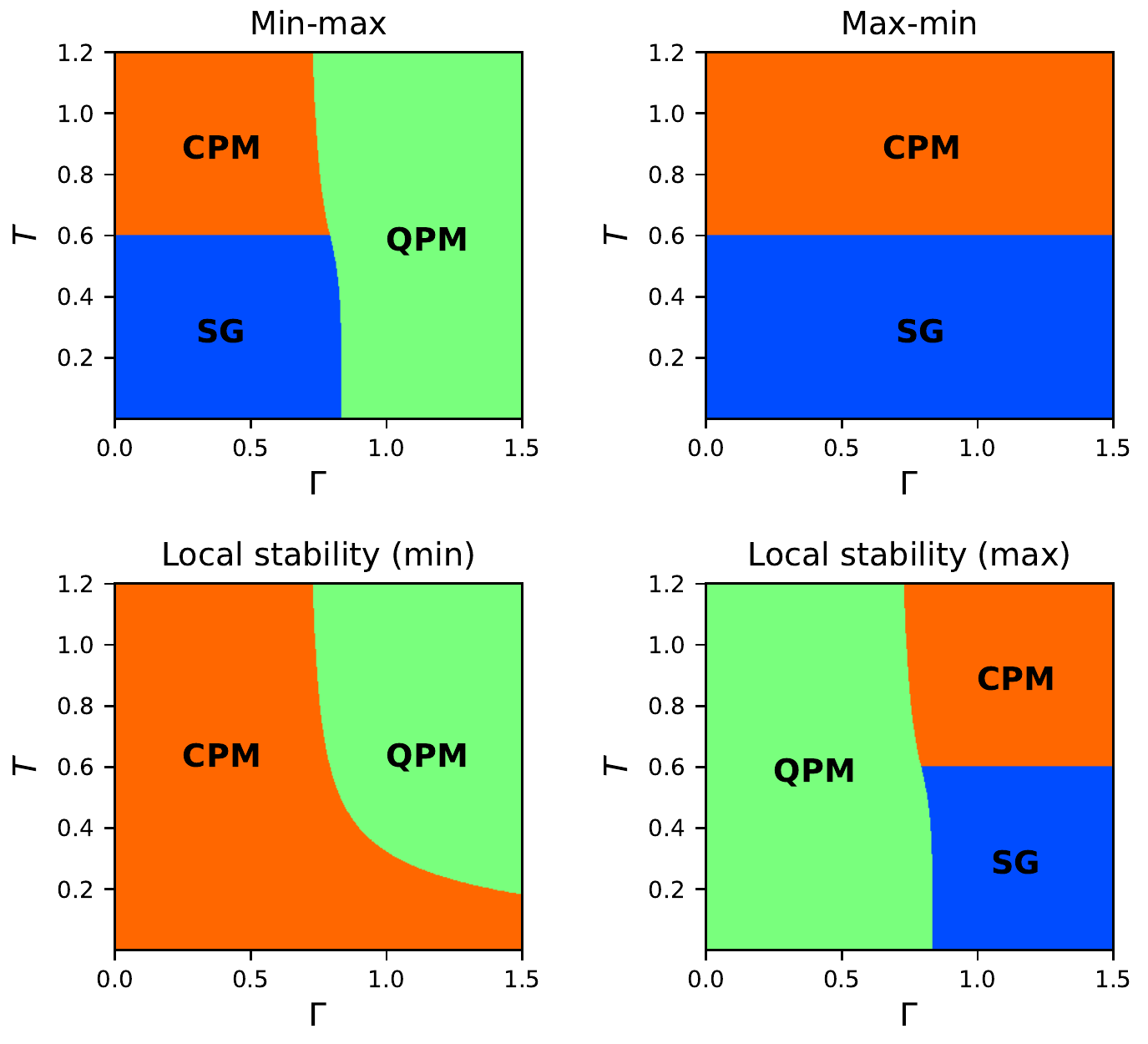}
\caption{Phase diagram for the transverse-field $p$-spin model (at large $p$) following different prescriptions for extremizing the action, as discussed in the main text. QPM stands for the ``quantum paramagnetic'' phase, CPM for ``classical paramagnetic'', and SG for ``spin glass''.}
\label{fig:possible_REM_phase_diagrams}
\end{figure}

Fig.~\ref{fig:possible_REM_phase_diagrams} shows the phase diagrams that result from many of these prescriptions.
The differences are rather striking.
The min-max phase diagram, first of all, is quite sensible and matches that reported in the literature~\cite{Goldschmidt1990Solvable,Nieuwenhuizen1998Quantum}.
The max-min phase diagram, however, completely lacks a QPM phase.
Minimizing among stable saddle points overlooks the entire SG phase, whereas maximizing gives inverted behavior as a function of $\Gamma$, predicting that the QPM phase would be dominant at low $\Gamma$ but not at high $\Gamma$.
Full minimization and full maximization are not shown, as they give particularly unrealistic results --- either would disregard all of the above saddle points for being local minima in the direction of $R$ but local maxima in the direction of $m$ and $Q$.
Thus it is clear that the issue of which prescription to follow does have significant consequences, and the purpose of the following sections is to provide evidence that Eq.~\eqref{eq:min_max_definition} is the correct one to follow.

Before proceeding, we also observe that at large $\Gamma$ and sufficiently low temperature, the system is indeed in a situation (using the min-max prescription) where $f_{\textrm{Q}} \neq f_{\textrm{A}}$ without there being any replica order --- $f_{\textrm{Q}}$ is given by the QPM saddle point and $f_{\textrm{A}}$ by CPM\footnote{One can easily confirm, by reproducing the steps from Eq.~\eqref{eq:p_spin_partition_function} to Eq.~\eqref{eq:p_spin_simplified_effective_action} but with $n = 1$ throughout, that calculating $f_{\textrm{A}}$ is indeed equivalent to minimizing over $R$ while fixing $m = 1$ and $Q = 0$, i.e., $f_{\textrm{A}} = \min_R S_n(R, 1, 0)$.}.
Furthermore, this occurs exactly as described in Sec.~\ref{subsec:summary_results}.
For values of $R$ near that of the CPM saddle point, the maximization over replica order parameters gives $m < 1$ and $Q \sim 1$, while for values of $R$ near that of QPM, instead $m = 1$ and $Q = 0$.

Lastly, it is interesting to consider the results of Refs.~\cite{Cugliandolo2000From,Cugliandolo2001Imaginary} from this perspective.
The authors analyzed the closely related quantum \textit{spherical} $p$-spin model, and in the course of determining the equilibrium phase diagram, observed that there can be multiple stable solutions to the saddle-point equations.
They used physical arguments to rule out certain solutions and obtain the correct phase diagram, but noticed that their reasoning does not correspond to any simple rule for comparing the free energies alone --- the higher free energy must be chosen when comparing certain solutions in certain parameter ranges, but the lower free energy must be chosen when comparing others.

Given the close analogy between the spherical model and the Ising model considered here, we expect the min-max prescription to provide an explanation for those observations.
The ``spurious'' paramagnetic solution found in Refs.~\cite{Cugliandolo2000From,Cugliandolo2001Imaginary} corresponds to the CPM saddle point here, and we indeed find that it is rendered invalid at low temperature by the maximization over spin glass order parameters.
We further find that the saddle point with the higher free energy is selected when crossing the continuous phase boundary, while that with the lower free energy (after neglecting the spurious solution) is selected when crossing the discontinuous boundary, all analogous to that of Refs.~\cite{Cugliandolo2000From,Cugliandolo2001Imaginary}.
It would be worthwhile to revisit the spherical model and explicitly confirm that the min-max prescription yields the correct behavior there --- we leave this for future work.

\section{Collections of random energy models} \label{sec:random_energy_models}

The random energy model (REM)~\cite{Derrida1980Random,Derrida1981Random,Gross1984Simplest} and its generalizations~\cite{Derrida1986Solution,Mottishaw1986First,Mora2008Random} occupy an important place in spin glass theory because they are some of very few models that can be solved straightforwardly both with and without the replica trick.
They can thus be used to test many of the assertions made in the replica theory.
In that spirit, here we consider a generalization of the REM that involves both conventional and replica order parameters, and for which different prescriptions for extremizing the action result in different free energies.
Only the min-max prescription leads to the correct free energy in all cases.

This same strategy was used in Ref.~\cite{Mottishaw1986First} for the special case of a spin-1 REM.
However, we disagree with the conclusion of that paper.
Ref.~\cite{Mottishaw1986First} claims that one should identify the set of ``locally stable'' saddle points, in the sense of being a local minimum with respect to conventional order parameters and a local maximum with respect to replica order parameters, and then select the locally stable saddle point with the lowest free energy.
This prescription strikes us as problematic for two reasons:
\begin{itemize}
    \item Many models exhibit first-order spin glass transitions, in which the replica order parameters jump discontinuously as one crosses the phase boundary~\cite{Gardner1985Spin,Gross1985Mean}.
    For those models that do not have any conventional order parameters, it is by now well-established that the saddle point with the \textit{highest} free energy still must be chosen, even though both saddle points in question remain locally stable across the transition.
    \item When a model does have both conventional and replica order parameters, the Hessian of the action will generically have off-diagonal terms which couple fluctuations in the two.
    Unless one is willing to disregard the off-diagonal elements (which we shall soon show is erroneous in any case), it is unclear even how to interpret the statement of local stability since the eigenvectors of the Hessian have components along both conventional and replica fluctuations\footnote{We did not run into this issue in Sec.~\ref{sec:p_spin} only because the off-diagonal elements of the Hessian happened to be subleading at large $p$. That said, we found that the prescription of minimizing among stable saddle points gave incorrect results nonetheless.}.
\end{itemize}
Thus we feel that the issue of which prescription to follow remains unsettled\footnote{There is another sense in which one could define local stability, in terms of the eigenvalues of the Hessian with respect to fluctuations in \textit{individual} elements of the overlap matrix $Q_{\alpha \alpha'}$. It is known that the dominant saddle point of the replicated action has \textit{positive} eigenvalues even in the $n \rightarrow 0$ limit~\cite{DeAlmeida1978Stability,Mezard2009}, just as a conventional action in terms of conventional order parameters would. We see no reason why the requirement of positive eigenvalues would not continue to hold for an action with simultaneous replica and conventional order parameters. However, this sense of local stability still does not allow one to choose from multiple locally stable saddle points at first-order transitions. Thus the issue of the correct prescription remains.}, and here consider an even broader class of REMs so as to help resolve it.

First we review the original REM.
It is a system of $N$ spin-1/2s for which the Hamiltonian $H_{\textrm{REM}}(\sigma^z)$ (diagonal in the $\sigma^z$ basis) is merely an independent Gaussian random variable for each of the $2^N$ basis states.
Each energy level has mean zero and variance $N/2$.
While technically a random variable, one can easily show (see Ref.~\cite{Mezard2009}) that with probability 1 in the $N \rightarrow \infty$ limit, the density of states at energy per spin $\epsilon$ scales as $\exp{[N(\log{2} - \epsilon^2)]}$ for $|\epsilon| \leq \sqrt{\log{2}}$ and is zero otherwise.
Thus the free energy per spin $f$ is, again with probability 1 in the $N \rightarrow \infty$ limit,
\begin{equation} \label{eq:REM_partition_function}
\begin{aligned}
f &= -\lim_{N \rightarrow \infty} \frac{1}{N \beta} \log{\int_{-\sqrt{\log{2}}}^{\sqrt{\log{2}}} \textrm{d}\epsilon \exp{\Big[ N \Big( \log{2} - \epsilon^2 - \beta \epsilon \Big) \Big]}} \\
&= \begin{cases} -\frac{1}{\beta} \log{2} - \frac{\beta}{4}, \; & \beta \leq 2 \sqrt{\log{2}} \\ -\sqrt{\log{2}}, \; & \beta > 2 \sqrt{\log{2}} \end{cases}.
\end{aligned}
\end{equation}
Note that since Eq.~\eqref{eq:REM_partition_function} holds with probability 1, it gives the average (quenched) free energy density as well.
Also note the phase transition at $\beta_c \equiv 2 \sqrt{\log{2}}$ --- the entropy is positive and the energy density varies with temperature for $\beta < \beta_c$, whereas the entropy is zero and the energy density is frozen at its lowest value for $\beta > \beta_c$.
The former is the ``paramagnetic'' phase and the latter is identified as the ``spin glass'' phase.

Our generalization, termed the ``multi-REM'', is simply a collection of REMs parametrized by the variable $R \in [0, 1]$.
At each value of $R$, we take there to be $\exp{[Ns(R)]}$ basis states.
Each of those $\exp{[Ns(R)]}$ energy levels is the sum of a deterministic term $Nh(R)$ and a REM term of variance $N \Delta(R)/2$ (again independent of all others).
The three functions $s(R)$, $h(R)$, and $\Delta(R)$ are parameters of the model.
For present purposes, the physical interpretation of the quantity $R$ is unimportant.
Note that the original REM is the special case
\begin{equation} \label{eq:multi_REM_original_limit}
s(R) = \begin{cases} \log{2}, \; & R = 1 \\ -\infty, \; & R < 1 \end{cases}, \qquad h(R) = 0, \qquad \Delta(R) = 1,
\end{equation}
and the spin-1 REM of Ref.~\cite{Mottishaw1986First} is the special case
\begin{equation} \label{eq:multi_REM_spin_1_limit}
s(R) = R \log{2} - R \log{R} - (1 - R) \log{(1 - R)}, \qquad h(R) = -DR, \qquad \Delta(R) = \begin{cases} 1, \; & R = 1 \\ 0, \; & R < 1 \end{cases}.
\end{equation}
Further examples, corresponding to the REM in a magnetic field or with additional ferromagnetic interactions, can be found in Ref.~\cite{Derrida1981Random}.

The exact same argument as for the REM implies that the density of states for the \textit{random term} at a given value of $R$ is $\exp{[N(s(R) - \epsilon^2/\Delta(R))]}$ for $|\epsilon| \leq \sqrt{\Delta(R) s(R)}$ and is zero otherwise.
Since the total energy is $Nh(R) + N \epsilon$, the free energy density is
\begin{equation} \label{eq:multi_REM_free_energy}
\begin{aligned}
f &= -\lim_{N \rightarrow \infty} \frac{1}{N \beta} \log{\int_0^1 \textrm{d}R \int_{-\sqrt{\Delta(R) s(R)}}^{\sqrt{\Delta(R) s(R)}} \textrm{d}\epsilon \exp{\left[ N \left( s(R) - \frac{\epsilon^2}{\Delta(R)} - \beta h(R) - \beta \epsilon \right) \right]}} \\
&= -\lim_{N \rightarrow \infty} \frac{1}{N \beta} \log{\int_0^1 \textrm{d}R \begin{cases} \exp{\Big[ N \Big( s(R) - \beta h(R) + \frac{\beta^2 \Delta(R)}{4} \Big) \Big]}, \; & \beta \leq 2 \sqrt{\frac{s(R)}{\Delta(R)}} \\ \exp{\Big[ N \Big( -\beta h(R) + \beta \sqrt{\Delta(R) s(R)} \Big) \Big]}, \; & \beta > 2 \sqrt{\frac{s(R)}{\Delta(R)}} \end{cases}} \\
&= \min_{R \in [0, 1]} \begin{cases} -\frac{s(R)}{\beta} + h(R) - \frac{\beta \Delta(R)}{4}, \; & \beta \leq 2 \sqrt{\frac{s(R)}{\Delta(R)}} \\ h(R) - \sqrt{\Delta(R) s(R)}, \; & \beta > 2 \sqrt{\frac{s(R)}{\Delta(R)}} \end{cases}.
\end{aligned}
\end{equation}
In cases where $s(R) = -\infty$ for certain values of $R$ (such as Eq.~\eqref{eq:multi_REM_original_limit} above), those values should simply be omitted from the final minimization over $R$.

Eq.~\eqref{eq:multi_REM_free_energy} is the exact expression for the free energy.
Let us now see how to recover it from the replica trick.
Denoting individual states by $\sigma$ and (somewhat sloppily) indicating those belonging to a given value of $R$ by $\sigma \in R$, the partition function is
\begin{equation} \label{eq:multi_REM_starting_partition_function}
Z = \int_0^1 \textrm{d}R \sum_{\sigma \in R} \exp{\big[ -N \beta h(R) - \beta E(\sigma) \big]},
\end{equation}
where $E(\sigma)$ is the REM contribution to the energy of state $\sigma$ --- it is a Gaussian of mean zero and variance $N \Delta(R)/2$, independent of all other energies.
The $n$'th moment of $Z$ is, after some algebra (see Ref.~\cite{Mezard2009}), 
\begin{equation} \label{eq:multi_REM_replicated_partition_function}
\mathbb{E}_E Z^n = \int_0^1 \prod_{\alpha = 1}^n \textrm{d}R_{\alpha} \sum_{\sigma^1 \in R_1} \cdots \sum_{\sigma^n \in R_n} \exp{\left[ -N \beta \sum_{\alpha = 1}^n h(R_{\alpha}) + \frac{N \beta^2}{4} \sum_{\alpha \alpha'} \delta_{\sigma^{\alpha} \sigma^{\alpha'}} \Delta(R_{\alpha}) \right]}.
\end{equation}
In REM-like models such as this, the factor $\delta_{\sigma^{\alpha} \sigma^{\alpha'}}$ serves as the overlap $Q_{\alpha \alpha'}$ --- it can only take the values 0 and 1, but we nonetheless extremize the action in Eq.~\eqref{eq:multi_REM_replicated_partition_function} over all possible values of the matrix $Q_{\alpha \alpha'}$.
Much as we did for the transverse-field $p$-spin model, assume that the dominant extremum has $R_{\alpha} = R$ independent of $\alpha$, and make a 1RSB ansatz --- the replicas divide into $n/m$ groups of $m$ each, with $\delta_{\sigma^{\alpha} \sigma^{\alpha'}} = 1$ (meaning $\sigma^{\alpha} = \sigma^{\alpha'}$) for $\alpha$ and $\alpha'$ in the same group, and $\delta_{\sigma^{\alpha} \sigma^{\alpha'}} = 0$ (meaning $\sigma^{\alpha} \neq \sigma^{\alpha'}$) otherwise.
The action of Eq.~\eqref{eq:multi_REM_replicated_partition_function} is then
\begin{equation} \label{eq:multi_REM_replicated_action}
S_n(R, m) \sim -\frac{s(R)}{m \beta} + h(R) - \frac{m \beta \Delta(R)}{4},
\end{equation}
and the question is once again how to select the dominant extremum.

First consider the min-max prescription.
Recalling that $m \in [0, 1]$ in the $n \rightarrow 0$ limit, we have that
\begin{equation} \label{eq:multi_REM_min_max_prescription_result}
\min_{R \in [0, 1]} \max_{m \in [0, 1]} S_n(R, m) = \min_{R \in [0, 1]} \begin{cases} -\frac{s(R)}{\beta} + h(R) - \frac{\beta \Delta(R)}{4}, \; & \beta \leq 2 \sqrt{\frac{s(R)}{\Delta(R)}} \\ h(R) - \sqrt{\Delta(R) s(R)}, \; & \beta > 2 \sqrt{\frac{s(R)}{\Delta(R)}} \end{cases}.
\end{equation}
This is identical to the exact expression in Eq.~\eqref{eq:multi_REM_free_energy}.
Thus the min-max prescription gives the correct result for any multi-REM, regardless of the forms of $s(R)$, $h(R)$, and $\Delta(R)$.

\begin{figure}[t]
\centering
\includegraphics[width=0.8\textwidth]{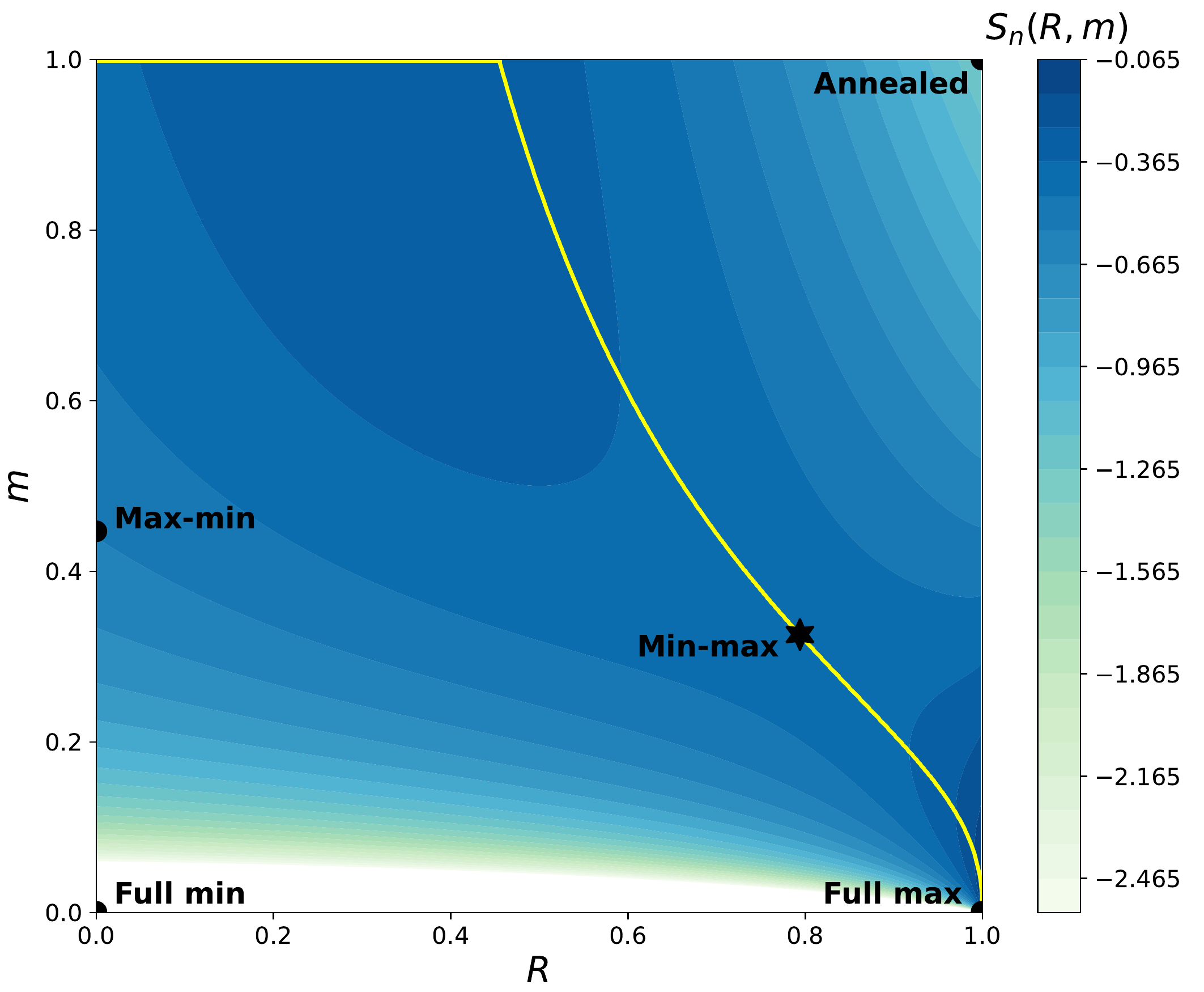}
\caption{Replicated action $S_n(R, m)$ (Eq.~\eqref{eq:multi_REM_replicated_action}) at $\beta = 5$ for the multi-REM with functions $s(R)$, $h(R)$, and $\Delta(R)$ given in Eq.~\eqref{eq:multi_REM_clarifying_example}. Sufficiently negative values of $S_n(R, m)$ are not colored (note that $S_n(R, m) \rightarrow -\infty$ as $m \rightarrow 0$ for all $R \neq 1$). Black symbols indicate the order parameter values obtained by extremizing $S_n(R, m)$ following different prescriptions --- min-max (star), max-min, full minimization, full maximization, annealed free energy (see Eqs.~\eqref{eq:min_max_definition} through~\eqref{eq:full_max_definition}). The yellow line is the curve $m(R)$ which maximizes $S_n(R, m)$ for each value of $R$ --- the min-max prescription corresponds to minimizing along this line.}
\label{fig:many_REM_landscape}
\end{figure}

It remains only to confirm that there exist instances of multi-REMs --- i.e., choices of $s(R)$, $h(R)$, and $\Delta(R)$ --- for which any other prescription gives a result different from Eq.~\eqref{eq:multi_REM_min_max_prescription_result}.
We can then claim that the min-max prescription is unambiguously the correct one.
Fig.~\ref{fig:many_REM_landscape} presents such an instance (or at least one for which the prescriptions of Sec.~\ref{sec:p_spin} all give distinct results), namely:
\begin{equation} \label{eq:multi_REM_clarifying_example}
s(R) = \frac{1+R}{2} \log{\frac{2}{1+R}} + \frac{1-R}{2} \log{\frac{2}{1-R}}, \qquad h(R) = -\frac{(1-R)^2}{4}, \qquad \Delta(R) = R^3.
\end{equation}
Furthermore, one can confirm that at the correct saddle point (indicated by the star in Fig.~\ref{fig:many_REM_landscape}), $\partial^2 S_n(R, m) / \partial R^2 < 0$.
This example demonstrates that even the local stability of the saddle point would be misleading without the min-max prescription in mind --- it appears to be unstable to fluctuations in $R$, but is stable along the path $(R, m(R))$ (solid line in Fig.~\ref{fig:many_REM_landscape}) due to the off-diagonal element of the Hessian.

\section{Generalized Aizenman-Sims-Starr scheme} \label{sec:Aizenman_Sims_Starr}

The Aizenman-Sims-Starr scheme~\cite{Aizenman2003Extended} is a variational expression for the free energy of the classical $p$-spin model.
It has played an essential role in the rigorous proofs of predictions from the replica theory~\cite{Bovier2006,Panchenko2013}.
Here we do not aim to obtain any rigorous results, but rather show how a straightforward generalization of the Aizenman-Sims-Starr scheme automatically identifies the min-max prescription as the correct procedure.
Special cases of this already exist in the mathematical literature~\cite{Panchenko2018FreeA,Panchenko2018FreeB,Camilli2022Inference}, even going further to obtain an explicit final expression for the free energy (which we do not do here).
First we review the original scheme following Ref.~\cite{Aizenman2003Extended}, and then present the min-max generalization.

\subsection{Existence of the free energy} \label{subsec:free_energy_existence}

Consider the classical $p$-spin model:
\begin{equation} \label{eq:classical_pS_definition}
H_J^{(N)}(\sigma) = \sum_{(i_1 \cdots i_p)} J_{i_1 \cdots i_p} \sigma_{i_1} \cdots \sigma_{i_p},
\end{equation}
where (unlike in Sec.~\ref{sec:p_spin}) each $\sigma_i$ is simply a classical variable taking values $\pm 1$ and $H_J^{(N)}(\sigma)$ is simply a function of the $2^N$ possible configurations $\sigma \equiv \{ \sigma_i \}_{i=1}^N$.
We indicate the system size as a superscript for later convenience.
For technical reasons that will become clear, we assume that $p$ is even.
The same results can be derived for odd $p$, but substantially more work is required.

The Aizenman-Sims-Starr scheme consists of two observations:
\begin{itemize}
    \item The free energy resulting from Eq.~\eqref{eq:classical_pS_definition} is equivalent to that of a model in which each spin interacts not with the other $N-1$ spins but with a ``bath'' of auxiliary spins.
    \item The free energy resulting from Eq.~\eqref{eq:classical_pS_definition} is greater than or equal to that of \textit{any} model in which each spin interacts with a ``bath'' having certain properties (including the bath alluded to in the previous point).
\end{itemize}

To derive the first statement, we begin by showing that the average free energy density $f$ satisfies
\begin{equation} \label{eq:free_energy_difference_expression}
f = \lim_{N \rightarrow \infty} \limsup_{L \rightarrow \infty} \frac{F^{(N+L)} - F^{(L)}}{N},
\end{equation}
where $F^{(N)}$ denotes the disorder-averaged free energy of size $N$.
Both the existence of a well-defined $f \equiv \lim_{N \rightarrow \infty} F^{(N)}/N$ in the first place and its equivalence to Eq.~\eqref{eq:free_energy_difference_expression} follow from the sub-additivity of the free energy, i.e., $F^{(N+L)} \leq F^{(N)} + F^{(L)}$ (see App.~\ref{app:subadditivity_consequences}).
Thus our first task amounts to proving this sub-additivity.

Let $\sigma \equiv \{ \sigma_i \}_{i=1}^N$ denote a set of $N$ spins and $\alpha \equiv \{ \alpha_j \}_{j=1}^L$ denote a separate set of $L$ spins.
Consider \textit{independent} Hamiltonians $H_J^{(N)}(\sigma)$, $H_J^{(L)}(\alpha)$, and $H_J^{(N+L)}(\sigma, \alpha)$.
All three are of the form in Eq.~\eqref{eq:classical_pS_definition}, but the couplings have different variances ($p!/2N^{p-1}$, $p!/2L^{p-1}$, and $p!/2(N+L)^{p-1}$ respectively).
In particular, the covariances of the energies are
\begin{equation} \label{eq:pS_covariances}
\begin{gathered}
\mathbb{E} H_J^{(N)}(\sigma) H_J^{(N)}(\sigma') \sim \frac{N}{2} \big( \sigma \cdot \sigma' \big)^p, \qquad \mathbb{E} H_J^{(L)}(\alpha) H_J^{(L)}(\alpha') \sim \frac{L}{2} \big( \alpha \cdot \alpha' \big)^p, \\
\mathbb{E} H_J^{(N+L)}(\sigma, \alpha) H_J^{(N+L)}(\sigma', \alpha') \sim \frac{N+L}{2} \left( \frac{N}{N+L} \big( \sigma \cdot \sigma' \big) + \frac{L}{N+L} \big( \alpha \cdot \alpha' \big) \right)^p,
\end{gathered}
\end{equation}
where we define
\begin{equation} \label{eq:pS_dot_product_definitions}
\sigma \cdot \sigma' \equiv \frac{1}{N} \sum_i \sigma_i \sigma'_i, \qquad \alpha \cdot \alpha' \equiv \frac{1}{L} \sum_j \alpha_j \alpha'_j.
\end{equation}
Lastly define the ``interpolation Hamiltonian''
\begin{equation} \label{eq:pS_subadditivity_interpolation_Hamiltonian}
H_J(\sigma, \alpha; \lambda) \equiv \sqrt{\lambda} \bigg( H_J^{(N)}(\sigma) + H_J^{(L)}(\alpha) \bigg) + \sqrt{1 - \lambda} \, H_J^{(N+L)}(\sigma, \alpha),
\end{equation}
and corresponding $\lambda$-dependent free energy $F(\lambda) \equiv -\beta^{-1} \mathbb{E} \log{\sum_{\sigma \alpha} \exp{[-\beta H_J(\sigma, \alpha; \lambda)]}}$.
Note that $F(0) = F^{(N+L)}$ and $F(1) = F^{(N)} + F^{(L)}$.
Thus we shall prove that $F^{(N+L)} \leq F^{(N)} + F^{(L)}$ by showing that $\textrm{d}F(\lambda)/\textrm{d}\lambda \geq 0$ for all $\lambda \in [0, 1]$.

The key tool is ``Gaussian integration by parts''~\cite{Talagrand2011a,Panchenko2013}: if $\{ X_a \}$ is any collection of mean-zero Gaussian random variables, and $f(X)$ is any differentiable function of the variables, then (assuming $f(X)$ does not grow so rapidly that the averages fail to exist)
\begin{equation} \label{eq:Gaussian_integration_by_parts}
\mathbb{E} \big[ X_a f(X) \big] = \sum_b \mathbb{E} \big[ X_a X_b \big] \mathbb{E} \left[ \frac{\partial f(X)}{\partial X_b} \right].
\end{equation}
We do not prove Eq.~\eqref{eq:Gaussian_integration_by_parts} here (see instead the above references), but note that the formula can be viewed as a generalization of Wick's theorem\footnote{To see this, consider the case where $f(X)$ is simply a product of Gaussians: $f(X) = \prod_k X_{b_k}$. Then Eq.~\eqref{eq:Gaussian_integration_by_parts} amounts to the statement $\mathbb{E} [X_a \prod_k X_{b_k}] = \sum_k \mathbb{E}[X_a X_{b_k}] \mathbb{E}[\prod_{l \neq k} X_{b_l}]$, which is Wick's theorem expressed recursively.}.
We shall repeatedly use Eq.~\eqref{eq:Gaussian_integration_by_parts} in situations analogous to the present one, i.e., where we have a Gaussian mean-zero Hamiltonian $H_J(\sigma, \alpha; \lambda)$ depending on a parameter $\lambda$, a ``bare'' probability distribution $w(\sigma, \alpha)$ (uniform in the present context but not more generally), the free energy
\begin{equation} \label{eq:Gaussian_integration_general_free_energy}
F(\lambda) \equiv -\frac{1}{\beta} \mathbb{E} \log{\sum_{\sigma \alpha} w(\sigma, \alpha) \exp{\Big[ -\beta H_J(\sigma, \alpha; \lambda) \Big]}},
\end{equation}
and we aim to compute $\textrm{d}F(\lambda)/\textrm{d}\lambda$.
Gaussian integration by parts gives
\begin{equation} \label{eq:free_energy_derivative_structure_derivation}
\begin{aligned}
\frac{\textrm{d}F(\lambda)}{\textrm{d}\lambda} &= \sum_{\sigma \alpha} \mathbb{E} \left[ \frac{\partial H_J(\sigma, \alpha; \lambda)}{\partial \lambda} \frac{w(\sigma, \alpha) \exp{\big[ -\beta H_J(\sigma, \alpha; \lambda) \big]}}{\sum_{\rho \gamma} w(\rho, \gamma) \exp{\big[ -\beta H_J(\rho, \gamma; \lambda) \big]}} \right] \\
&= \sum_{\sigma \alpha} \sum_{\sigma' \alpha'} \mathbb{E} \left[ \frac{\partial H_J(\sigma, \alpha; \lambda)}{\partial \lambda} H_J(\sigma', \alpha'; \lambda) \right] \\
&\qquad \qquad \qquad \cdot \mathbb{E} \left[ \frac{\partial}{\partial H_J(\sigma', \alpha'; \lambda)} \frac{w(\sigma, \alpha) \exp{\big[ -\beta H_J(\sigma, \alpha; \lambda) \big]}}{\sum_{\rho \gamma} w(\rho, \gamma) \exp{\big[ -\beta H_J(\rho, \gamma; \lambda) \big]}} \right] \\
&= -\beta \sum_{\sigma \alpha} \sum_{\sigma' \alpha'} \mathbb{E} \left[ \frac{\partial H_J(\sigma, \alpha; \lambda)}{\partial \lambda} H_J(\sigma', \alpha'; \lambda) \right] \\
&\qquad \qquad \qquad \cdot \mathbb{E} \left[ \delta_{\sigma \sigma'} \delta_{\alpha \alpha'} \frac{w(\sigma, \alpha) \exp{\big[ -\beta H_J(\sigma, \alpha; \lambda) \big]}}{\sum_{\rho \gamma} w(\rho, \gamma) \exp{\big[ -\beta H_J(\rho, \gamma; \lambda) \big]}} \right. \\
&\qquad \qquad \qquad \qquad \qquad \left. - \frac{w(\sigma, \alpha) w(\sigma', \alpha') \exp{\big[ -\beta H_J(\sigma, \alpha; \lambda) - \beta H_J(\sigma', \alpha'; \lambda) \big]}}{\big( \sum_{\rho \gamma} w(\rho, \gamma) \exp{\big[ -\beta H_J(\rho, \gamma; \lambda) \big]} \big)^2} \right].
\end{aligned}
\end{equation}
Denote by $\langle \, \cdot \, \rangle_{\lambda}^{(1)}$ the ``one-replica'' thermal expectation value
\begin{equation} \label{eq:one_replica_average_expectation_value}
\big< \cdots \big>_{\lambda}^{(1)} \equiv \sum_{\sigma \alpha} \cdots \mathbb{E} \frac{w(\sigma, \alpha) \exp{\big[ -\beta H_J(\sigma, \alpha; \lambda) \big]}}{\sum_{\rho \gamma} w(\rho, \gamma) \exp{\big[ -\beta H_J(\rho, \gamma; \lambda) \big]}},
\end{equation}
and denote by $\langle \, \cdot \, \rangle_{\lambda}^{(2)}$ the ``two-replica'' thermal expectation value
\begin{equation} \label{eq:two_replica_average_expectation_value}
\big< \cdots \big>_{\lambda}^{(2)} \equiv \sum_{\sigma \alpha} \sum_{\sigma' \alpha'} \cdots \mathbb{E} \frac{w(\sigma, \alpha) w(\sigma', \alpha') \exp{\big[ -\beta H_J(\sigma, \alpha; \lambda) - \beta H_J(\sigma', \alpha'; \lambda) \big]}}{\big( \sum_{\rho \gamma} w(\rho, \gamma) \exp{\big[ -\beta H_J(\rho, \gamma; \lambda) \big]} \big)^2}.
\end{equation}
Then we can express Eq.~\eqref{eq:free_energy_derivative_structure_derivation} succinctly as
\begin{equation} \label{eq:free_energy_derivative_structure}
\frac{1}{\beta} \frac{\textrm{d}F(\lambda)}{\textrm{d}\lambda} = -\Big< \mathbb{E} \left[ \frac{\partial H_J(\sigma, \alpha; \lambda)}{\partial \lambda} H_J(\sigma, \alpha; \lambda) \right] \Big>_{\lambda}^{(1)} + \Big< \mathbb{E} \left[ \frac{\partial H_J(\sigma, \alpha; \lambda)}{\partial \lambda} H_J(\sigma', \alpha'; \lambda) \right] \Big>_{\lambda}^{(2)}.
\end{equation}
Note that Eq.~\eqref{eq:free_energy_derivative_structure} holds regardless of the degrees of freedom being summed over (even the decomposition into $\sigma$ and $\alpha$ is unnecessary) and regardless of the bare distribution $w(\sigma, \alpha)$.

In the present case, using Eq.~\eqref{eq:pS_subadditivity_interpolation_Hamiltonian} with Eq.~\eqref{eq:pS_covariances},
\begin{equation} \label{eq:pS_subadditivity_derivative_correlations}
\begin{aligned}
\mathbb{E} \left[ \frac{\partial H_J(\sigma, \alpha; \lambda)}{\partial \lambda} H_J(\sigma', \alpha'; \lambda) \right] &= \frac{N+L}{4} \left[ \frac{N}{N+L} \big( \sigma \cdot \sigma' \big)^p + \frac{L}{N+L} \big( \alpha \cdot \alpha' \big)^p \right. \\
&\qquad \qquad \qquad \qquad \left. - \left( \frac{N}{N+L} \big( \sigma \cdot \sigma' \big) + \frac{L}{N+L} \big( \alpha \cdot \alpha' \big) \right)^p \right].
\end{aligned}
\end{equation}
The right-hand side has the property (at least for even $p$) that it is automatically non-negative and vanishes when $\sigma = \sigma'$ and $\alpha = \alpha'$ --- the non-negativity because the function $x^p$ is convex, and the vanishing because $\sigma \cdot \sigma = \alpha \cdot \alpha = 1$.
Thus the one-replica term of Eq.~\eqref{eq:free_energy_derivative_structure} is zero and the two-replica term is non-negative.
This proves that $\textrm{d}F(\lambda)/\textrm{d}\lambda \geq 0$ and $F^{(N+L)} \leq F^{(N)} + F^{(L)}$, as claimed.

\subsection{The original variational principle} \label{subsec:original_Aizenman_Sims_Starr}

With Eq.~\eqref{eq:free_energy_difference_expression} in hand, we now use the indicated order of limits ($1 \ll N \ll L$) to simplify $F^{(N+L)}$ and $F^{(L)}$.
For the size-$(N+L)$ system, again denote the first $N$ spins by $\{ \sigma_i \}_{i=1}^N$ and the remaining $L$ spins by $\{ \alpha_j \}_{j=1}^{L}$ , and write the Hamiltonian as
\begin{equation} \label{eq:p_spin_system_bath_decomposition}
H_J^{(N+L)}(\sigma, \alpha) = E_{(0,p)}(\alpha) + E_{(1, p-1)}(\sigma, \alpha) + \cdots + E_{(p, 0)}(\sigma),
\end{equation}
where $E_{(q,p-q)}(\sigma, \alpha)$ denotes the sum of terms involving $q$ of the $N$ spins and $p - q$ of the $L$ spins:
\begin{equation} \label{eq:p_spin_system_bath_subterm}
E_{(q,p-q)}(\sigma, \alpha) \equiv \sum_{(i_1 \cdots i_q)} \sum_{(j_{q+1} \cdots j_p)} J_{i_1 \cdots i_q j_{q+1} \cdots j_p} \sigma_{i_1} \cdots \sigma_{i_q} \alpha_{j_{q+1}} \cdots \alpha_{j_p}.
\end{equation}
Note that here each coupling has variance $p!/2(N+L)^{p-1}$, and thus $E_{(q,p-q)}(\sigma, \alpha)$ has variance scaling with $N$ and $L$ as $N^q L^{1-q}$ (times a prefactor).
One can show (see Ref.~\cite{Aizenman2003Extended}) that the terms with $q \geq 2$ do not affect the free energy, since their variances vanish in the limit $N/L \ll 1$.
Thus for the purpose of evaluating $f$, we can approximate the size-$(N+L)$ Hamiltonian by
\begin{equation} \label{eq:p_spin_system_bath_Hamiltonian}
\begin{aligned}
H_J^{(N+L)}(\sigma, \alpha) &\sim E_{(0,p)}(\alpha) + E_{(1, p-1)}(\sigma, \alpha) \\
&= E_{(0,p)}(\alpha) + \sum_i \left( \sum_{(j_2 \cdots j_p)} J_{i j_2 \cdots j_p} \alpha_{j_2} \cdots \alpha_{j_p} \right) \sigma_i.
\end{aligned}
\end{equation}

As for the size-$L$ system, its Hamiltonian is not quite equal to $E_{(0,p)}(\alpha)$ because the couplings in the latter have slightly too small variance.
However, we do have the equality (in distribution)
\begin{equation} \label{eq:p_spin_bath_Hamiltonian}
H_J^{(L)}(\alpha) = E_{(0,p)}(\alpha) + \sum_{(j_1 \cdots j_p)} J'_{j_1 \cdots j_p} \alpha_{j_1} \cdots \alpha_{j_p},
\end{equation}
where $J'$ is independent of $E_{(0,p)}$ and
\begin{equation} \label{eq:p_spin_bath_Hamiltonian_correction_variance}
\mathbb{E} J_{j_1 \cdots j_p}'^2 = \frac{p!}{2} \left( \frac{1}{L^{p-1}} - \frac{1}{(N+L)^{p-1}} \right) \sim \frac{(p-1) N}{2} \frac{p!}{L^p}.
\end{equation}
Thus define the quantities
\begin{equation} \label{eq:p_spin_bath_effective_terms_definition}
h_i(\alpha) \equiv \sum_{(j_2 \cdots j_p)} J_{i j_2 \cdots j_p} \alpha_{j_2} \cdots \alpha_{j_p}, \qquad U(\alpha) \equiv \sum_{(j_1 \cdots j_p)} J'_{j_1 \cdots j_p} \alpha_{j_1} \cdots \alpha_{j_p},
\end{equation}
which are independent of each other and have covariances (in the limit $1 \ll N \ll L$)
\begin{equation} \label{eq:p_spin_bath_effective_terms_covariance}
\mathbb{E} h_i(\alpha) h_{i'}(\alpha') \sim \delta_{ii'} \frac{p}{2} \big( \alpha \cdot \alpha' \big)^{p-1}, \qquad \mathbb{E} U(\alpha) U(\alpha') \sim \frac{(p-1) N}{2} \big( \alpha \cdot \alpha' \big)^p.
\end{equation}
Note that we are again using the dot product defined in Eq.~\eqref{eq:pS_dot_product_definitions}.
Also define the probability distribution
\begin{equation} \label{eq:p_spin_bath_bare_distribution}
w(\alpha) \equiv \frac{\exp{\left[ -\beta E_{(0,p)}(\alpha) \right]}}{\sum_{\gamma} \exp{\left[ -\beta E_{(0,p)}(\gamma) \right]}}.
\end{equation}
Eqs.~\eqref{eq:p_spin_system_bath_Hamiltonian} and~\eqref{eq:p_spin_bath_Hamiltonian} together with these definitions allow us to write Eq.~\eqref{eq:free_energy_difference_expression} as (with the limit $1 \ll N \ll L$ implied)
\begin{equation} \label{eq:p_spin_free_energy_Aizenman_Sims_Starr}
f \sim -\frac{1}{N \beta} \mathbb{E} \log{\sum_{\sigma \alpha} w(\alpha) \exp{\left[ -\beta \sum_i h_i(\alpha) \sigma_i \right]}} + \frac{1}{N \beta} \mathbb{E} \log{\sum_{\alpha} w(\alpha) \exp{\Big[ -\beta U(\alpha) \Big]}}.
\end{equation}
The averages in Eq.~\eqref{eq:p_spin_free_energy_Aizenman_Sims_Starr} are over all random variables --- $E_{(0,p)}$, $h_i$, and $U$.

Eq.~\eqref{eq:p_spin_free_energy_Aizenman_Sims_Starr} can indeed be interpreted as each spin $\sigma_i$ experiencing only a local field $h_i(\alpha)$ depending on ``bath'' degrees of freedom labeled by $\alpha$ (plus a correction term depending solely on the bath).
The factor $w(\alpha)$ is simply a Boltzmann distribution for the bath states (see Eq.~\eqref{eq:p_spin_bath_bare_distribution}).
While not useful on its own, since the right-hand side is no easier to evaluate than the original free energy, this expression does suggest the form that a more general bath should take if to be compared against the $p$-spin model.
Miraculously, as we explain in what follows, the right-hand side of Eq.~\eqref{eq:p_spin_free_energy_Aizenman_Sims_Starr} evaluated for any such bath is less than or equal to $f$, which gives us the variational expression we seek.

Thus now let $\alpha$ denote \textit{any} degrees of freedom and $w(\alpha)$ denote \textit{any} probability distribution on $\alpha$.
Suppose there is an associated dot product $\alpha \cdot \alpha'$ with $\alpha \cdot \alpha = 1$.
Define the Gaussian random functions $h_i(\alpha)$ and $U(\alpha)$ (independent of each other) such that
\begin{equation} \label{eq:ROSt_fields_covariance}
\mathbb{E} h_i(\alpha) h_{i'}(\alpha') = \delta_{ii'} \frac{p}{2} \big( \alpha \cdot \alpha' \big)^{p-1}, \qquad \mathbb{E} U(\alpha) U(\alpha') = \frac{(p-1) N}{2} \big( \alpha \cdot \alpha' \big)^p,
\end{equation}
and lastly define
\begin{equation} \label{eq:ROSt_action_definition}
\overline{f} \equiv -\frac{1}{N \beta} \mathbb{E} \log{\sum_{\sigma \alpha} w(\alpha) \exp{\left[ -\beta \sum_i h_i(\alpha) \sigma_i \right]}} + \frac{1}{N \beta} \mathbb{E} \log{\sum_{\alpha} w(\alpha) \exp{\Big[ -\beta U(\alpha) \Big]}},
\end{equation}
where the averages\footnote{One can allow for $w(\alpha)$ to itself be random and independent of the other quantities, as is the case in Eq.~\eqref{eq:p_spin_bath_bare_distribution}. Since $f \geq \overline{f}$ for any individual realization of $w(\alpha)$, it trivially holds that $f \geq \mathbb{E}_w \overline{f}$ as well.} are over $h_i$ and $U$.
Note that we recover the previous expressions by taking $\alpha$ to be the set of $L$ spins in a size-$(N+L)$ $p$-spin model, with $\alpha \cdot \alpha' \equiv L^{-1} \sum_j \alpha_j \alpha'_j$ and $w(\alpha)$ given by Eq.~\eqref{eq:p_spin_bath_bare_distribution}.
Ref.~\cite{Aizenman2003Extended} refers to any such bath as a ``random overlap structure'' (ROSt) --- we have that the $L$ spins of a $p$-spin model themselves form a ROSt.

To show that $f \geq \overline{f}$, regardless of any further properties of the bath, define another ``interpolation Hamiltonian''
\begin{equation} \label{eq:p_spin_interpolation_Hamiltonian}
H_J(\sigma, \alpha; \lambda) \equiv \sqrt{\lambda} \Big( H_J^{(N)}(\sigma) + U(\alpha) \Big) + \sqrt{1 - \lambda} \sum_i h_i(\alpha) \sigma_i,
\end{equation}
with free energy $g(\lambda) \equiv -(N \beta)^{-1} \mathbb{E} \log{\sum_{\sigma \alpha} w(\alpha) \exp{[-\beta H_J(\sigma, \alpha; \lambda)]}}$.
Note that $g(1) - g(0) = f - \overline{f}$.
Thus the statement $f \geq \overline{f}$ is equivalent to $g(1) \geq g(0)$.
The derivative $\textrm{d}g(\lambda)/\textrm{d}\lambda$ is again given by Eq.~\eqref{eq:free_energy_derivative_structure} (up to a factor of $N$), and we have that
\begin{equation} \label{eq:p_spin_derivative_correlations}
\mathbb{E} \left[ \frac{\partial H_J(\sigma, \alpha; \lambda)}{\partial \lambda} H_J(\sigma', \alpha'; \lambda) \right] = \frac{N}{4} \bigg[ \big( \sigma \cdot \sigma' \big)^p + (p-1) \big( \alpha \cdot \alpha' \big)^p - p \big( \alpha \cdot \alpha' \big)^{p-1} \big( \sigma \cdot \sigma' \big) \bigg].
\end{equation}
Once again, the right-hand side is (for even $p$) non-negative due to convexity and vanishing when $\sigma = \sigma'$ and $\alpha = \alpha'$, meaning $\textrm{d}g(\lambda)/\textrm{d}\lambda \geq 0$ and $f \geq \overline{f}$.

To summarize, this proves that the free energy of the $p$-spin model is greater than or equal to the ``free energy'' (defined in Eq.~\eqref{eq:ROSt_action_definition}) of any ROSt.
At the same time, we know there is at least one ROSt (the $L$ spins of a larger $p$-spin model) whose free energy equals that of the $p$-spin model.
We can thus somewhat schematically write
\begin{equation} \label{eq:Aizenman_Sims_Starr_variational_principle}
f = \max_{\textrm{ROSt}} \overline{f}(\textrm{ROSt}),
\end{equation}
i.e., the free energy is obtained by maximizing $\overline{f}$ over all possible ROSts.
This is the Aizenman-Sims-Starr variational expression.
Note that it quite suggestively involves a \textit{maximization} rather than minimization over ROSts.

Admittedly, Eq.~\eqref{eq:Aizenman_Sims_Starr_variational_principle} is still not a useful expression on its own, since the task of evaluating $\overline{f}$ for all possible ROSts appears hopelessly difficult.
Yet there turns out to be a subset of ROSts for which $\overline{f}$ can be evaluated explicitly, and furthermore, the resulting expression agrees exactly with the action emerging from the replica theory~\cite{Panchenko2013}.
It can also be shown (through much work) that no other ROSt can give a larger $\overline{f}$ than those in this subset.
Thus this line of reasoning, which we shall not expand on any further (see instead Refs.~\cite{Bovier2006,Talagrand2011a,Talagrand2011b,Panchenko2013}), ultimately provides a rigorous justification for the claim that one must maximize the replicated action to obtain the correct free energy of the $p$-spin model.

\subsection{The generalized variational principle --- naive attempt} \label{subsec:generalized_Aizenman_Sims_Starr_naive}

The preceding subsection was concerned entirely with the classical $p$-spin model.
We now show how to generalize to more complicated models, in particular many which involve conventional order parameters.
As a sufficiently broad class of models, let $\vec{\sigma}$ denote any set of degrees of freedom indexed by both a ``spin'' label $i \in \{1, \cdots, N\}$ and ``component'' label $\tau \in \{1, \cdots, M\}$.
An individual element of $\vec{\sigma}$ is written $\sigma_i(\tau)$.
The set of all components of a given spin is denoted $\vec{\sigma}_i \equiv \{ \sigma_i(\tau) \}_{\tau}$, and the set of all $\tau$-components is denoted $\sigma(\tau) \equiv \{ \sigma_i(\tau) \}_i$.
Each individual $\sigma_i(\tau)$ is a classical but otherwise arbitrary variable, and there can be constraints between the different components of each $\vec{\sigma}_i$.

We take the Hamiltonian to be\footnote{Here the sum over $(i_1 \cdots i_p)$ is over all tuples of $p$ spin indices, i.e., over all $i_1$ through $i_p$ such that $i_1 < \cdots < i_p$, whereas the sum over $\tau_1 \cdots \tau_p$ is over all $\tau_1$ through $\tau_p$ without any restriction on the ordering.}
\begin{equation} \label{eq:fully_generic_Hamiltonian}
\begin{aligned}
H^{(N)}(\vec{\sigma}) &= \sum_{(i_1 \cdots i_p)} \sum_{\tau_1 \cdots \tau_p} J_{i_1 \cdots i_p}^{\tau_1 \cdots \tau_p} \sigma_{i_1}(\tau_1) \cdots \sigma_{i_p}(\tau_p) \, + \, \sum_i H_0(\vec{\sigma}_i) \\
&\equiv H_J^{(N)}(\vec{\sigma}) + \sum_i H_0(\vec{\sigma}_i),
\end{aligned}
\end{equation}
where $H_0$ is an arbitrary single-site term, and where the couplings are mean-zero Gaussians that are independent with respect to \textit{spin} indices but can have almost arbitrary correlations with respect to \textit{component} indices (including restrictions on which combinations of components are allowed):
\begin{equation} \label{eq:fully_generic_couplings}
\mathbb{E} \left[ J_{i_1 \cdots i_p}^{\tau_1 \cdots \tau_p} J_{i_1' \cdots i_p'}^{\tau_1' \cdots \tau_p'} \right] = \frac{p!}{2N^{p-1}} \delta_{i_1 i_1'} \cdots \delta_{i_p i_p'} C_{\tau_1 \cdots \tau_p}^{\tau_1' \cdots \tau_p'}.
\end{equation}

Our analysis requires only two conditions on the correlation matrix $C$.
The first is simply that it should be permutation-symmetric: for any permutation $\pi$ of the numbers 1 through $p$,
\begin{equation} \label{eq:generic_coupling_correlation_condition_1}
C_{\tau_1 \cdots \tau_p}^{\tau_1' \cdots \tau_p'} = C_{\tau_{\pi(1)} \cdots \tau_{\pi(p)}}^{\tau_{\pi(1)}' \cdots \tau_{\pi(p)}'}.
\end{equation}
Note that the same permutation $\pi$ enters for both primed and unprimed indices.
Eq.~\eqref{eq:generic_coupling_correlation_condition_1} should be considered as merely part of the definition that the Hamiltonian is ``mean-field'', with statistically equivalent interactions between all sets of spins.
As a result of it, the covariance structure of the energies takes a relatively simple form:
\begin{equation} \label{eq:fully_generic_energy_covariances}
\mathbb{E} H_J^{(N)}(\vec{\sigma}) H_J^{(N)}(\vec{\sigma}') \sim \frac{N}{2} V(\sigma \cdot \sigma'),
\end{equation}
\begin{equation} \label{eq:fully_generic_correlation_function_definition}
V(\sigma \cdot \sigma') \equiv \sum_{\tau_1 \cdots \tau_p} \sum_{\tau_1' \cdots \tau_p'} C_{\tau_1 \cdots \tau_p}^{\tau_1' \cdots \tau_p'} \big[ \sigma(\tau_1) \cdot \sigma'(\tau_1') \big] \cdots \big[ \sigma(\tau_p) \cdot \sigma'(\tau_p') \big],
\end{equation}
where the matrix of dot products $\sigma \cdot \sigma' \equiv \{ \sigma(\tau) \cdot \sigma'(\tau') \}_{\tau \tau'}$ is defined as
\begin{equation} \label{eq:fully_generic_dot_products}
\sigma(\tau) \cdot \sigma'(\tau') \equiv \frac{1}{N} \sum_i \sigma_i(\tau) \sigma_i'(\tau').
\end{equation}

The second condition on $C$ is that the function $V(\sigma \cdot \sigma')$, viewed as a function on the space of $M \times M$ matrices, is convex\footnote{To be completely explicit, we are requiring that for any matrices $X$ and $Y$, and any $\lambda \in [0, 1]$, \begin{equation*} V \big( (1 - \lambda) X + \lambda Y \big) \leq (1 - \lambda) V(X) + \lambda V(Y). \end{equation*} An immediate consequence is that, again for any $X$ and $Y$, \begin{equation*} V(X) + \sum_{\tau \tau'} Y_{\tau \tau'} \frac{\partial V(X)}{\partial X_{\tau \tau'}} \leq V(X + Y). \end{equation*}}.
This is a technical assumption needed for the proof, analogous to how we took $p$ to be even in the preceding subsections.
We expect our conclusions regarding the min-max prescription to hold more generally (although see Ref.~\cite{Mourrat2021Nonconvex}).

This class of models strikes us as the most general to share the all-to-all random interaction structure of the $p$-spin model.
There are numerous examples of independent interest:
\begin{itemize}
    \item \textbf{Original $p$-spin model}: $M = 1$, with $\sigma_i \in \{+1, -1\}$.
    Note that we now allow for longitudinal fields by the inclusion of the single-site term $H_0$.
    \item \textbf{Higher spins}: Again $M = 1$, but now $\sigma_i$ takes values in an arbitrary set $\{s_1, \cdots, s_S\}$.
    \item \textbf{Spherical models}: Still $M = 1$, now with $\sigma_i \in (-\infty, \infty)$.
    To keep the spectrum bounded, one usually imposes the ``spherical constraint'' $\sum_i \sigma_i^2 = N$.
    While technically outside the class of models we are considering, these can easily be included by adding a Lagrange multiplier to $H_0$ that enforces the spherical constraint~\cite{Crisanti1992Spherical,Crisanti1993Spherical}.
    \item \textbf{Classical rotors}: There are many ways to generalize the $p$-spin model to higher-component spins.
    Take each $\sigma_i(\tau) \in [-S, S]$ for some $S > 0$, with the constraint that $\sum_{\tau} \sigma_i(\tau)^2 = S^2$ for each $i$.
    Our flexibility in choosing $C$ allows for many different types of interactions even within this situation.
    \item \textbf{Transverse-field Ising models (in Suzuki-Trotter representation)}: Large $M$, with $\sigma_i(\tau) \in \{+1, -1\}$.
    $H_0$ should include a term coming from the transverse field under Trotterization, as in Sec.~\ref{sec:p_spin}.
    However, keep in mind that to truly relate to a quantum transverse-field model, one would require the opposite limit ($M \rightarrow \infty$ at finite $N$) to that considered here (finite $M$ as $N \rightarrow \infty$).
    \item \textbf{Quantum-mechanical particles}: Again large $M$, now with $\sigma_i(\tau) \in (-\infty, \infty)$ subject to a spherical constraint.
    Compared to transverse-field Ising models, one only needs a different $H_0$ coming instead from the Trotterization of the kinetic energy $\hat{P}_i^2$.
    The subtlety about taking $M \rightarrow \infty$ vs $N \rightarrow \infty$ still applies.
    \item \textbf{Coherent-state path integrals for spins}: Depending on the interaction structure between quantum spins, it may be more convenient to use coherent states as the basis for the path integral~\cite{Auerbach1994}.
    This still fits into the class of models we consider, but slightly more thought is required --- $\tau$ should label both imaginary time and the spin component, with constraints among those spin components at the same imaginary time.
    Regardless, $H_0$ still includes a term coming from Trotterization, and the $M \rightarrow \infty$ vs $N \rightarrow \infty$ comment still applies.
\end{itemize}
Our analysis of the general model in Eq.~\eqref{eq:fully_generic_Hamiltonian} immediately applies to all of these situations.
While many of them do not involve literal spins, we shall continue to refer to each $\vec{\sigma}_i$ as a spin and each $\sigma_i(\tau)$ as a component of that spin.

We first attempt to apply the analysis of Secs.~\ref{subsec:free_energy_existence} and~\ref{subsec:original_Aizenman_Sims_Starr}.
We shall immediately run into difficulties, even for simply proving the sub-additivity of the free energy.
The remedy that we propose will in fact allow us to carry out all subsequent steps of the analysis as well, and in doing so, the min-max prescription will emerge naturally.

As before, consider spins $\{ \vec{\sigma}_i \}_{i=1}^N$ and $\{ \vec{\alpha}_j \}_{j=1}^L$, together with the three independent Hamiltonians $H^{(N)}(\vec{\sigma})$, $H^{(L)}(\vec{\alpha})$, and $H^{(N+L)}(\vec{\sigma}, \vec{\alpha})$.
The random terms of the Hamiltonians have covariances given by Eq.~\eqref{eq:fully_generic_energy_covariances} and the analogous expressions for the size-$L$ and size-$(N+L)$ systems.
Define the interpolation Hamiltonian
\begin{equation} \label{eq:subadditivity_attempt_interpolation_Hamiltonian}
\begin{aligned}
H(\vec{\sigma}, \vec{\alpha}; \lambda) &\equiv \sqrt{\lambda} \bigg( H_J^{(N)}(\vec{\sigma}) + H_J^{(L)}(\vec{\alpha}) \bigg) + \sqrt{1 - \lambda} \, H_J^{(N+L)}(\vec{\sigma}, \vec{\alpha}) + \sum_i H_0(\vec{\sigma}_i) + \sum_j H_0(\vec{\alpha}_j) \\
&\equiv H_J(\vec{\sigma}, \vec{\alpha}; \lambda) + \sum_i H_0(\vec{\sigma}_i) + \sum_j H_0(\vec{\alpha}_j),
\end{aligned}
\end{equation}
with free energy $F(\lambda) \equiv -\beta^{-1} \mathbb{E} \log{\sum_{\vec{\sigma} \vec{\alpha}} \exp{[-\beta H(\vec{\sigma}, \vec{\alpha}; \lambda)]}}$.
Note that it is only the random part of the Hamiltonians that we interpolate between as $\lambda$ varies.
We still have that $F(0) = F^{(N+L)}$ and $F(1) = F^{(N)} + F^{(L)}$, so sub-additivity would again result from showing that $\textrm{d}F(\lambda) / \textrm{d}\lambda \geq 0$.
The derivative is given by Eq.~\eqref{eq:free_energy_derivative_structure} (with the single-site terms as part of the bare distribution $w(\vec{\sigma}, \vec{\alpha})$), and we now have that
\begin{equation} \label{eq:naive_generic_derivative_correlations}
\begin{aligned}
\mathbb{E} \left[ \frac{\partial H_J(\vec{\sigma}, \vec{\alpha}; \lambda)}{\partial \lambda} H_J(\vec{\sigma}', \vec{\alpha}'; \lambda) \right] &= \frac{N+L}{4} \left[ \frac{N}{N+L} V(\sigma \cdot \sigma') + \frac{L}{N+L} V(\alpha \cdot \alpha') \right. \\
&\qquad \qquad \qquad \qquad \left. - V \left( \frac{N}{N+L} \sigma \cdot \sigma' + \frac{L}{N+L} \alpha \cdot \alpha' \right) \right].
\end{aligned}
\end{equation}
Here we run into a problem --- although Eq.~\eqref{eq:naive_generic_derivative_correlations} is again automatically non-negative due to the convexity of $V$, the model under consideration need \textit{not} have $\sigma \cdot \sigma = \alpha \cdot \alpha$ for all $\vec{\sigma}$ and $\vec{\alpha}$, even for single-component spins\footnote{As a simple example, consider a spin-1 model: $\sigma_i \in \{-1, 0, 1\}$. Then $\sigma \cdot \sigma \equiv N^{-1} \sum_i \sigma_i^2$ can lie anywhere between 0 and 1 depending on the configuration.}.
Thus the one-replica term of Eq.~\eqref{eq:free_energy_derivative_structure} need \textit{not} be zero, and since the one- and two-replica terms come with opposite signs, there is no reason to expect $\textrm{d}F(\lambda)/\textrm{d}\lambda \geq 0$ for all (or any) $\lambda \in [0, 1]$.

\subsection{The generalized variational principle --- min-max prescription} \label{subsec:generalized_Aizenman_Sims_Starr_min_max}

We circumvent the above issue by restricting the trace to be only over configurations having certain values of $\sigma(\tau) \cdot \sigma(\tau')$ and $\alpha(\tau) \cdot \alpha(\tau')$.
Namely, pick some symmetric matrix $R_{\tau \tau'}$ and define the $R$-dependent ``restricted partition function''
\begin{equation} \label{eq:R_dependent_partition_function}
Z^{(N)}(R) \equiv {\sum_{\vec{\sigma}}}^{(R)} \exp{\big[ -\beta H^{(N)}(\vec{\sigma}) \big]},
\end{equation}
where the superscript $(R)$ indicates that only those configurations with $\sigma(\tau) \cdot \sigma(\tau') = R_{\tau \tau'}$ for all $\tau$ and $\tau'$ (abbreviated $\sigma \cdot \sigma = R$) are to be summed over.
The full partition function can then be written as the sum of $Z^{(N)}(R)$ over all possible values of $R$:
\begin{equation} \label{eq:full_partition_function_decomposition}
Z^{(N)} = \sum_R Z^{(N)}(R).
\end{equation}
Define $f(R) \equiv -\lim_{N \rightarrow \infty} (N \beta)^{-1} \mathbb{E} \log{Z^{(N)}(R)}$ and $f \equiv -\lim_{N \rightarrow \infty} (N \beta)^{-1} \mathbb{E} \log{Z^{(N)}}$.
We shall derive an Aizenman-Sims-Starr variational expression for each $f(R)$ (compare to Eq.~\eqref{eq:Aizenman_Sims_Starr_variational_principle}): \begin{equation} \label{eq:R_dependent_Aizenman_Sims_Starr_variational_principle}
f(R) = \max_{\textrm{ROSt}} \overline{f}(R, \textrm{ROSt}).
\end{equation}
Then Eq.~\eqref{eq:full_partition_function_decomposition} can be evaluated by saddle-point at large $N$ to give
\begin{equation} \label{eq:generalized_Aizenman_Sims_Starr_variational_princple}
f = \min_R \max_{\textrm{ROSt}} \overline{f}(R, \textrm{ROSt}).
\end{equation}
Note that this is precisely the min-max prescription, interpreting the matrix $R_{\tau \tau'}$ as a set of conventional order parameters\footnote{Again considering the spin-1 example, $R$ is the value of $N^{-1} \sum_i \sigma_i^2$, which is precisely the order parameter used in Ref.~\cite{Mottishaw1986First} to analyze the model at large $p$.} --- after all, the need to introduce $R_{\tau \tau'}$ stems solely from the structure of the configuration space, without any reference to replicas or correlations between energy levels.

In the replica theory, $R_{\tau \tau'}$ appears as the ``self-overlap'', i.e., the value of the diagonal entries of the overlap matrix.
While simply 1 in the classical $p$-spin model, they can vary and must be integrated over more generally.
The transverse-field $p$-spin model of Sec.~\ref{sec:p_spin} provides an explicit example --- in the course of simplifying the $n$'th moment of the partition function (Eq.~\eqref{eq:p_spin_partition_function_evaluation_v2}), we found the need to introduce not only the inter-replica overlap $Q_{\alpha \alpha'}(\tau, \tau') \equiv N^{-1} \sum_i \sigma_i^{\alpha}(\tau) \sigma_i^{\alpha'}(\tau')$ but also the intra-replica overlap $R_{\alpha}(\tau, \tau') \equiv N^{-1} \sum_i \sigma_i^{\alpha}(\tau) \sigma_i^{\alpha}(\tau')$.
As argued then and confirmed now by Eq.~\eqref{eq:generalized_Aizenman_Sims_Starr_variational_princple}, the minimization over the self-overlap $R$ must be performed after the maximization over $Q$.

To prove Eqs.~\eqref{eq:R_dependent_Aizenman_Sims_Starr_variational_principle} and~\eqref{eq:generalized_Aizenman_Sims_Starr_variational_princple}, we start with an analogue to Eq.~\eqref{eq:free_energy_difference_expression}:
\begin{equation} \label{eq:generalized_free_energy_difference_expression}
f(R) = \lim_{N \rightarrow \infty} \limsup_{L \rightarrow \infty} \frac{F^{(N,L)}(R,R) - F^{(L)}(R)}{N},
\end{equation}
where $F^{(N,L)}(R,R)$ is the free energy of a size-$(N+L)$ system with \textit{separate} restrictions $\sigma \cdot \sigma = R$ and $\alpha \cdot \alpha = R$ (note that this is stricter than simply requiring the total self-overlap be $R$).
Eq.~\eqref{eq:generalized_free_energy_difference_expression} follows from an analogue of sub-additivity: $F^{(N,L)}(R,R) \leq F^{(N)}(R) + F^{(L)}(R)$ (see App.~\ref{app:subadditivity_consequences}).
Thus we first prove this inequality.

Define the interpolation Hamiltonian $H(\vec{\sigma}, \vec{\alpha}; \lambda)$ exactly as in Eq.~\eqref{eq:subadditivity_attempt_interpolation_Hamiltonian} (hence we still have Eq.~\eqref{eq:naive_generic_derivative_correlations} as well), but now with the interpolation free energy
\begin{equation} \label{eq:subadditivity_success_interpolation_free_energy}
F(R, R; \lambda) \equiv -\frac{1}{\beta} \mathbb{E} \log{{\sum_{\vec{\sigma} \vec{\alpha}}}^{(R,R)} \exp{\big[ -\beta H(\vec{\sigma}, \vec{\alpha}; \lambda) \big]}}.
\end{equation}
where $(R,R)$ indicates that the sums are only over $\vec{\sigma}$ and $\vec{\alpha}$ with $\sigma \cdot \sigma = \alpha \cdot \alpha = R$.
Thus in applying Eq.~\eqref{eq:free_energy_derivative_structure} with Eq.~\eqref{eq:naive_generic_derivative_correlations}, the one-replica term does now vanish --- only states with $\sigma \cdot \sigma = \alpha \cdot \alpha = R$ enter into the thermal expectation values to begin with.
The derivative $\partial F(R, R; \lambda) / \partial \lambda$ is non-negative, and $F(R, R; 0) = F^{(N,L)}(R, R)$ is less than or equal to $F(R, R; 1) = F^{(N)}(R) + F^{(L)}(R)$.
Eq.~\eqref{eq:generalized_free_energy_difference_expression} follows.

Since we are taking $M$ finite as $N, L \rightarrow \infty$ (in the order $1 \ll N \ll L$), the power-counting of Sec.~\ref{subsec:original_Aizenman_Sims_Starr} continues to apply here.
Thus
\begin{equation} \label{eq:fully_generic_system_bath_Hamiltonian}
H^{(N+L)}(\vec{\sigma}, \vec{\alpha}) \sim E_{(0,p)}(\vec{\alpha}) + \sum_{i \tau} h_i(\vec{\alpha}; \tau) \sigma_i(\tau) + \sum_i H_0(\vec{\sigma}_i) + \sum_j H_0(\vec{\alpha}_j),
\end{equation}
\begin{equation} \label{eq:fully_generic_bath_Hamiltonian}
H^{(L)}(\vec{\alpha}) \sim E_{(0,p)}(\vec{\alpha}) + U(\vec{\alpha}) + \sum_j H_0(\vec{\alpha}_j),
\end{equation}
where $h_i(\vec{\alpha}; \tau)$ and $U(\vec{\alpha})$ are straightforward generalizations of the expressions in Eq.~\eqref{eq:p_spin_bath_effective_terms_definition}, with covariances
\begin{equation} \label{eq:fully_generic_effective_field_term_covariance}
\begin{aligned}
\mathbb{E} h_i(\vec{\alpha}; \tau) h_{i'}(\vec{\alpha}'; \tau') &= \delta_{ii'} \frac{p}{2} \sum_{\tau_2 \cdots \tau_p} \sum_{\tau_2' \cdots \tau_p'} C_{\tau \tau_2 \cdots \tau_p}^{\tau' \tau_2' \cdots \tau_p'} \big[ \alpha(\tau_2) \cdot \alpha'(\tau_2') \big] \cdots \big[ \alpha(\tau_p) \cdot \alpha'(\tau_p') \big] \\
&= \delta_{ii'} \frac{1}{2} \frac{\partial V(\alpha \cdot \alpha')}{\partial [\alpha(\tau) \cdot \alpha'(\tau')]},
\end{aligned}
\end{equation}
\begin{equation} \label{eq:fully_generic_effective_correction_term_covariance}
\begin{aligned}
\mathbb{E} U(\vec{\alpha}) U(\vec{\alpha}') &= \frac{(p-1)N}{2} V(\alpha \cdot \alpha') \\
&= \frac{N}{2} \left[ \sum_{\tau \tau'} \big[ \alpha(\tau) \cdot \alpha'(\tau') \big] \frac{\partial V(\alpha \cdot \alpha')}{\partial [\alpha(\tau) \cdot \alpha'(\tau')]} - V(\alpha \cdot \alpha') \right].
\end{aligned}
\end{equation}
The second line of Eq.~\eqref{eq:fully_generic_effective_field_term_covariance} uses the permutation symmetry of $C$, and the second line of Eq.~\eqref{eq:fully_generic_effective_correction_term_covariance} uses that $\sum_{\tau \tau'} X_{\tau \tau'} \partial V(X) / \partial X_{\tau \tau'} = p V(X)$.
Define
\begin{equation} \label{eq:fully_generic_bare_bath_probability_distribution}
w(\vec{\alpha}) \equiv \frac{\exp{\left[ -\beta E_{(0,p)}(\vec{\alpha}) - \beta \sum_j H_0(\vec{\alpha}_j) \right]}}{\sum_{\vec{\gamma}}^{(R)} \exp{\left[ -\beta E_{(0,p)}(\vec{\gamma}) - \beta \sum_j H_0(\vec{\gamma}_j) \right]}},
\end{equation}
and we can then express $f(R)$ as
\begin{equation} \label{eq:fully_generic_free_energy_Aizenman_Sims_Starr_form}
\begin{aligned}
f(R) &= -\frac{1}{N \beta} \mathbb{E} \log{{\sum_{\vec{\sigma} \vec{\alpha}}}^{(R,R)} w(\vec{\alpha}) \exp{\left[ -\beta \sum_{i \tau} h_i(\vec{\alpha}; \tau) \sigma_i(\tau) - \beta \sum_i H_0(\vec{\sigma}_i) \right]}} \\
&\qquad \qquad \qquad \qquad + \frac{1}{N \beta} \mathbb{E} \log{{\sum_{\vec{\alpha}}}^{(R)} w(\vec{\alpha}) \exp{\Big[ -\beta U(\vec{\alpha}) \Big]}}.
\end{aligned}
\end{equation}

The last step is to consider more general baths, still with a fixed value of $R$, and show that $f(R) \geq \overline{f}(R)$ for the corresponding bath ``free energy'' $\overline{f}(R)$.
Apart from the presence of $R$, this is exactly analogous to what was done in Sec.~\ref{subsec:original_Aizenman_Sims_Starr}.
Thus let $\vec{\alpha}$ now denote any degrees of freedom, still labeled by $\tau$ but not necessarily by $i$.
Let $w(\vec{\alpha})$ denote any probability distribution on $\vec{\alpha}$, and let $\alpha(\tau) \cdot \alpha'(\tau')$ denote any dot product \textit{such that} $\alpha(\tau) \cdot \alpha(\tau') = R_{\tau \tau'}$.
Define Gaussian random functions $h_i(\vec{\alpha}; \tau)$ and $U(\vec{\alpha})$ with the same covariance structure as in Eqs.~\eqref{eq:fully_generic_effective_field_term_covariance} and~\eqref{eq:fully_generic_effective_correction_term_covariance}, and define
\begin{equation} \label{eq:fully_generic_ROSt_action_definition}
\begin{aligned}
\overline{f}(R) &\equiv -\frac{1}{N \beta} \mathbb{E} \log{{\sum_{\vec{\sigma} \vec{\alpha}}}^{(R,R)} w(\vec{\alpha}) \exp{\left[ -\beta \sum_{i \tau} h_i(\vec{\alpha}; \tau) \sigma_i(\tau) - \beta \sum_i H_0(\vec{\sigma}_i) \right]}} \\
&\qquad \qquad \qquad \qquad + \frac{1}{N \beta} \mathbb{E} \log{{\sum_{\vec{\alpha}}}^{(R)} w(\vec{\alpha}) \exp{\Big[ -\beta U(\vec{\alpha}) \Big]}}.
\end{aligned}
\end{equation}
We use the interpolation technique one final time.
Define
\begin{equation} \label{eq:fully_generic_bound_interpolation_Hamiltonian}
\begin{aligned}
H(\vec{\sigma}, \vec{\alpha}; \lambda) &\equiv \sqrt{\lambda} \bigg( H_J^{(N)}(\vec{\sigma}) + U(\vec{\alpha}) \bigg) + \sqrt{1 - \lambda} \sum_{i \tau} h_i(\vec{\alpha}; \tau) \sigma_i(\tau) + \sum_i H_0(\vec{\sigma}_i) \\
&\equiv H_J(\vec{\sigma}, \vec{\alpha}; \lambda) + \sum_i H_0(\vec{\sigma}_i),
\end{aligned}
\end{equation}
with
\begin{equation} \label{eq:fully_generic_bound_interpolation_free_energy}
g(R; \lambda) \equiv -\frac{1}{N \beta} \mathbb{E} \log{{\sum_{\vec{\sigma} \vec{\alpha}}}^{(R,R)} w(\vec{\alpha}) \exp{\big[ -\beta H(\vec{\sigma}, \vec{\alpha}; \lambda) \big]}},
\end{equation}
so that $g(R; 1) - g(R; 0) = f(R) - \overline{f}(R)$.
Using Eq.~\eqref{eq:free_energy_derivative_structure} with
\begin{equation} \label{eq:fully_generic_bound_derivative_correlations}
\begin{aligned}
\mathbb{E} \left[ \frac{\partial H_J(\vec{\sigma}, \vec{\alpha}; \lambda)}{\partial \lambda} H_J(\vec{\sigma}', \vec{\alpha}'; \lambda) \right] &= \frac{N}{4} \left[ \vphantom{\sum_{\tau \tau'}} V(\sigma \cdot \sigma') - V(\alpha \cdot \alpha') \right. \\
&\qquad \qquad \left. - \sum_{\tau \tau'} \Big[ \sigma(\tau) \cdot \sigma'(\tau') - \alpha(\tau) \cdot \alpha'(\tau') \Big] \frac{\partial V(\alpha \cdot \alpha')}{\partial [\alpha(\tau) \cdot \alpha'(\tau')]} \right],
\end{aligned}
\end{equation}
the restriction to states with $\sigma \cdot \sigma = \alpha \cdot \alpha = R$ again ensures that the one-replica term vanishes.
Thus $\partial g(R; \lambda) / \partial \lambda \geq 0$, and $f(R) \geq \overline{f}(R)$.

To reiterate, we have shown that the \textit{restricted} free energy $f(R)$ can be expressed as the maximum of $\overline{f}(R, \textrm{ROSt})$ over all ROSts having the same value of the self-overlap $R$.
Since $f$ is (almost by definition) the minimum of $f(R)$, this establishes the min-max prescription in Eq.~\eqref{eq:generalized_Aizenman_Sims_Starr_variational_princple}.

However, keep in mind that this does not constitute a proof of the min-max prescription, since we have not carried out the remaining (much harder) steps needed to verify the replica results --- proving that the replicated effective action is equivalent to $\overline{f}(R, \textrm{ROSt})$ for a tractable subset of ROSts, and then proving that the global maximum is attained among that subset.
It very well may be that subsequent steps cannot be carried out so straightforwardly simply by considering a restricted free energy, although complete proofs do already exist for certain special cases~\cite{Panchenko2018FreeA,Panchenko2018FreeB,Camilli2022Inference}.
Further investigation is certainly warranted.

\section{Conclusion} \label{sec:conclusion}

We have shown that when applying the replica trick to a model with not only spin glass order but additional types as well, the correct procedure is to first maximize the effective action with respect to replica order parameters $Q$ and then minimize with respect to the remaining ``conventional'' order parameters $R$.
As a result, one should consider the question of spin glass order (or replica order more generally) \textit{separately} for each value of $R$ --- there can be (and in fact often is) spin glass order for certain values of $R$ but not for others.
Whether the equilibrium state of the system has spin glass order depends on which value of $R$ gives the lowest free energy.

This distinction is especially important in regimes where the tendency for spin glass order competes with other types of order (such as at low temperature and high field in the transverse-field $p$-spin model).
In that case, we have shown that different prescriptions for applying the replica trick can lead to dramatically different phase diagrams (see Fig.~\ref{fig:possible_REM_phase_diagrams}).

The min-max prescription advocated for here also sheds light on the relationship between the quenched and annealed free energies.
While it is straightforward to see that the two free energies can differ without any spin glass order (the SK model in a longitudinal field provides a simple example), the fact that there need not be any replica order whatsoever is more subtle, since the actions being extremized to calculate the two become identical when $Q = 0$.
Yet according to the min-max prescription, the fact that $f_{\textrm{Q}} \neq f_{\textrm{A}}$ only implies that there is replica order for some value $R_{\textrm{A}}$ which may not be the equilibrium value $R_{\textrm{Q}}$.
In this sense, quite general arguments showing that $f_{\textrm{Q}} \neq f_{\textrm{A}}$ at low temperature in mean-field spin models (see Ref.~\cite{Baldwin2020Quenched}) in fact imply very little about the quenched system itself, at least on their own.

An important question going forward is the extent to which these conclusions apply beyond mean-field theory.
The analysis presented here is limited to models with infinite-range Gaussian random interactions --- the free energy reduces to an extremization over a matrix $Q_{\alpha \alpha'}$ and vector $R_{\alpha}$ only in such cases.
While these models are already quite interesting and important, it would certainly be worthwhile to investigate whether there are any implications to the quenched and annealed free energies agreeing more generally.
We leave this for future work.

\appendix

\section{Extremizing with respect to order parameters vs Lagrange multipliers} \label{app:conventional_extremizing}

Entirely unrelated to the replica trick and spin glass physics, there are subtleties in how one extremizes over (conventional) order parameters and their associated Lagrange multipliers.
This is an old topic and we are certainly not the first to consider it (to the point where it often passes without comment in the literature).
Yet since the present work is specifically concerned with the order in which one extremizes an action with respect to various quantities, we feel that it is appropriate to give a clear discussion of the issue here.

As a concrete example, consider a classical spin-1/2 Ising model with ``mean-field'' interactions:
\begin{equation} \label{eq:simple_Ising_mean_field_Hamiltonian}
H(\sigma) = N \epsilon \bigg( N^{-1} \sum_i \sigma_i \bigg),
\end{equation}
for some function $\epsilon(m)$.
In other words, the energy can be written as a function solely of the magnetization density $N^{-1} \sum_i \sigma_i$.
To evaluate the partition function, we can separate the trace into an outer sum over values of the magnetization $m$ and an inner sum over $\sigma$ such that $N^{-1} \sum_i \sigma_i = m$:
\begin{equation} \label{eq:simple_Ising_partition_decomposition}
Z \equiv \sum_{\sigma} \exp{\big[ -\beta H(\sigma) \big]} = \int_{-1}^1 \textrm{d}m \exp{\big[ -N \beta \epsilon(m) \big]} \sum_{\sigma} \delta \bigg( m - N^{-1} \sum_i \sigma_i \bigg).
\end{equation}
Defining $\sum_{\sigma} \delta(m - N^{-1} \sum_i \sigma_i) \equiv \exp{[Ns(m)]}$ and evaluating the integral over $m$ by saddle point, we have that
\begin{equation} \label{eq:simple_Ising_partition_evaluation}
-\lim_{N \rightarrow \infty} (N \beta)^{-1} \log{Z} = \min_{m \in [-1, 1]} \Big[ \epsilon(m) - \beta^{-1} s(m) \Big].
\end{equation}

Let us pretend that we do not have an explicit expression for $\exp{[Ns(m)]}$ --- while it is simply a binomial coefficient in the present example, it may not have a closed form more generally.
There are then two ways to proceed.
One often sees the $\delta$-function expressed in integral form as $(2\pi)^{-1} \int_{-i \infty}^{i \infty} N \textrm{d}h \exp{\big[ -Nhm + h \sum_i \sigma_i \big]}$ (note that $h$ runs along the imaginary axis).
We will discuss this approach momentarily.
Alternatively, one can use a method more along the lines of large deviation theory~\cite{denHollander2000} and consider the auxiliary quantity
\begin{equation} \label{eq:simple_Ising_noninteracting_problem}
Z_0(h) \equiv \sum_{\sigma} \exp{\Big[ \beta h \sum_i \sigma_i \Big]} = \exp{\big[ -N \beta g(h) \big]},
\end{equation}
where $g(h) \equiv -\beta^{-1} \log{2 \cosh{\beta h}}$.
Since one could again separate the sum over $\sigma$ into an outer and inner sum just as in Eq.~\eqref{eq:simple_Ising_partition_decomposition}, we have that\footnote{Note that Eq.~\eqref{eq:simple_Ising_noninteracting_relationship} establishes $g(h)$ as the Legendre transform of $s(m)$. The discussion that follows is really just an explanation of how to invert the Legendre transform.}
\begin{equation} \label{eq:simple_Ising_noninteracting_relationship}
g(h) = \min_{m \in [-1, 1]} \Big[ -hm - \beta^{-1} s(m) \Big].
\end{equation}
Denote the location of the minimum, which will be a function of $h$, by $m^*(h)$.
There is an explicit expression for $m^*(h)$: just as Eq.~\eqref{eq:simple_Ising_noninteracting_problem} is dominated by $\sigma$ with magnetizations close to $m^*(h)$, so is
\begin{equation} \label{eq:simple_Ising_noninteracting_magnetization_expression}
\frac{\partial \log{Z_0(h)}}{\partial h} = \sum_{\sigma} \bigg( \beta \sum_i \sigma_i \bigg) \frac{\exp{\big[ \beta h \sum_i \sigma_i \big]}}{\sum_{\sigma'} \exp{\big[ \beta h \sum_i \sigma'_i \big]}} \sim N \beta m^*(h),
\end{equation}
i.e., $m^*(h) = -\partial g(h) / \partial h$.
At this point, note that if one chooses $h$ so that\footnote{Since $\partial m^*(h)/\partial h$ is always positive (as one can explicitly check from Eq.~\eqref{eq:simple_Ising_noninteracting_magnetization_expression}) and $\lim_{h \rightarrow \pm \infty} m^*(h) = \pm 1$, there is exactly one solution to $m^*(h) = m$ for all $m \in (-1, 1)$.} $m^*(h) = m$, then from Eq.~\eqref{eq:simple_Ising_noninteracting_relationship}, one has $\beta^{-1} s(m) = -hm - g(h)$ and
\begin{equation} \label{eq:simple_Ising_partition_actual_evaluation}
-\lim_{N \rightarrow \infty} (N \beta)^{-1} \log{Z} = \min_{m \in [-1, 1]} \Big[ \epsilon(m) + hm + g(h) \Big].
\end{equation}
Since we have an explicit expression\footnote{One might wonder why we allow ourselves to use the explicit expression for $g(h)$ when we are pretending to not know the result for $s(m)$. Generically, evaluating $s(m)$ directly will involve a sum over all $N$ degrees of freedom subject to a constraint (here that $\sum_i \sigma_i = Nm$). On the other hand, $Z_0(h)$ is a non-interacting partition function, and thus evaluating $g(h)$ involves a single sum over one degree of freedom. The latter is often significantly simpler, hence the reason to consider $g(h)$.} for $g(h)$, Eq.~\eqref{eq:simple_Ising_partition_actual_evaluation} can readily be evaluated (keeping in mind that $h$ is a function of $m$ defined by $m^*(h) = m$).

In fact, since $h$ solves the equation $m = m^*(h) = -\partial g(h) / \partial h$, we can view $h$ as being determined by extremizing the ``action'' $\epsilon(m) + hm + g(h)$ at fixed $m$.
Thus the free energy is determined by extremizing with respect to both $m$ and $h$.
However, note that the second derivative with respect to $h$ is
\begin{equation} \label{eq:simple_Ising_noninteracting_second_derivative}
\frac{\partial^2 g(h)}{\partial h^2} = -N \beta \left[ \Big< \bigg( N^{-1} \sum_i \sigma_i \bigg)^2 \Big> - \Big< N^{-1} \sum_i \sigma_i \Big>^2 \right],
\end{equation}
where $\langle \, \cdot \, \rangle$ denotes a thermal expectation value with respect to $h \sum_i \sigma_i$.
Thus the second derivative is automatically negative, and the free energy is \textit{maximized} with respect to $h$.
Since $h$ is really a function of $m$ in Eq.~\eqref{eq:simple_Ising_partition_actual_evaluation}, the maximization occurs inside the minimization, meaning we can write
\begin{equation} \label{eq:simple_Ising_partition_final_evaluation}
-\lim_{N \rightarrow \infty} (N \beta)^{-1} \log{Z} = \min_{m \in [-1, 1]} \max_h \Big[ \epsilon(m) + hm + g(h) \Big].
\end{equation}
Interestingly, this is another ``min-max'' prescription, albeit one unrelated to that of the main text (although see Ref.~\cite{Barbier2019Adaptive} for an alternate derivation using the interpolation techniques of Sec.~\ref{sec:Aizenman_Sims_Starr}).

The same min-max prescription is hidden within the approach to calculating $s(m)$ based on the integral representation of $\delta(m - N^{-1} \sum_i \sigma_i)$.
In this approach, starting from Eq.~\eqref{eq:simple_Ising_partition_decomposition}, we have that
\begin{equation} \label{eq:simple_Ising_partition_integral_representation}
Z \sim \int_{-1}^1 \textrm{d}m \int_{-i \infty}^{i \infty} \textrm{d}h \exp{\Big[ -N \beta \big[ \epsilon(m) + hm + g(h) \big] \Big]},
\end{equation}
with the same $g(h)$ as defined in Eq.~\eqref{eq:simple_Ising_noninteracting_problem}.
The right-hand side can be evaluated by saddle point, but since $h$ initially runs along the imaginary axis, its contour must be deformed to pass through the (real) solution to $m = -\partial g(h) / \partial h$.
We do need that the action be minimized with respect to $h$ \textit{along the trajectory of the contour}, but this is fully consistent with the fact that $\partial^2 g(h) / \partial h^2 < 0$ for real $h$ since the contour passes through the solution vertically.
The second derivative being negative in the real direction implies that it is positive in the imaginary direction, as required.
Thus we are in fact maximizing the action with respect to real $h$ after all.

Regardless of the approach, it is clear that $m$ and $h$ play different roles.
$m$ is undeniably an order parameter --- from the beginning, we use it to decompose the original partition function (Eq.~\eqref{eq:simple_Ising_partition_decomposition}).
$h$ is instead a Lagrange multiplier --- we use it to enforce the constraint that $N^{-1} \sum_i \sigma_i = m$.
The conclusion here is that one should first maximize the effective action with respect to Lagrange multipliers, and then minimize with respect to order parameters.

With the proper ordering in mind, let us lastly consider the solution to the saddle point equations.
We have the pair
\begin{equation} \label{eq:simple_Ising_saddle_point_equations}
\frac{\partial g(h)}{\partial h} = -m, \qquad \frac{\partial \epsilon(m)}{\partial m} = -h,
\end{equation}
where the former is to be solved for $h$, and then the latter is to be solved for $m$.
Nonetheless, it is tempting to interpret the latter equation as determining $h$ and then use that expression in the former to obtain an equation for $m$.
In fact, this is what we do in Sec.~\ref{sec:p_spin} of the main text --- we use Eq.~\eqref{eq:p_spin_easy_saddle_points} to solve for the \textit{Lagrange multipliers} $K$ and $\Lambda$.
Although decidedly \textit{not} the procedure we have derived thus far, the substitution $h = -\partial \epsilon(m) / \partial m$ turns out to be justified, as we now show.

To be precise, let $h^*(m)$ be the solution to $\partial g(h) / \partial h = -m$, and let $h^{\times}(m)$ denote the function $-\partial \epsilon(m) / \partial m$.
We have already established that the correct free energy is obtained by minimizing $S^*(m) \equiv \epsilon(m) + h^* m + g(h^*)$ with respect to $m$ --- taking a derivative (assuming the minimum lies in the interior\footnote{This is natural to expect --- $\partial S^*(m) / \partial m = \partial \epsilon(m) / \partial m + h^*(m)$ and $h^*(m) \rightarrow \pm \infty$ as $m \rightarrow \pm 1$, meaning the minimum cannot lie at either endpoint unless $\partial \epsilon(m) / \partial m$ diverges there.} of $[-1, 1]$) leads to the equation $\partial \epsilon(m) / \partial m = -h^*(m)$.
Now instead consider minimizing $S^{\times}(m) \equiv \epsilon(m) + h^{\times} m + g(h^{\times})$ --- taking a derivative gives $[m + \partial g(h^{\times}) / \partial h^{\times}] \partial h^{\times} / \partial m = 0$.
Thus unless $\partial h^{\times} / \partial m = 0$ (a case that can often be treated separately\footnote{For example, suppose $\epsilon(m) = m^p$ for $p > 2$. Then $\partial h^{\times}/\partial m$ does equal 0 at $m = 0$, but this is a stationary point of $S^*(m)$ anyway.}), the extrema of $S^{\times}$ occur where $m = -\partial g(h^{\times}) / \partial h^{\times}$.
Either way --- whether minimizing $S^*$ or $S^{\times}$ --- the same equations are being solved (namely Eq.~\eqref{eq:simple_Ising_saddle_point_equations}) and the same two-parameter action is being evaluated (namely $\epsilon(m) + hm + g(h)$).
Thus the correct global minimum is identified (except for points at which $\partial h^{\times} / \partial m = 0$).
This is true even though $S^*(m) \neq S^{\times}(m)$ for general values of $m$.

\section{Consequences of sub-additivity} \label{app:subadditivity_consequences}

We demonstrated in Sec.~\ref{subsec:free_energy_existence} of the main text that the disorder-averaged free energy of the classical $p$-spin model is sub-additive, $F^{(N+L)} \leq F^{(N)} + F^{(L)}$ (where the superscript indicates the system size).
This implies both that $f \equiv \lim_{N \rightarrow \infty} F^{(N)}/N$ exists and that it can be written as in Eq.~\eqref{eq:free_energy_difference_expression}, reproduced here:
\begin{equation} \label{eq:free_energy_difference_expression_reproduced}
f = \lim_{N \rightarrow \infty} \limsup_{L \rightarrow \infty} \frac{F^{(N+L)} - F^{(L)}}{N}.
\end{equation}
For completeness, we prove this statement here (following Ref.~\cite{Panchenko2013}).

The fact that sub-additivity implies the existence of $\lim_{N \rightarrow \infty} F^{(N)}/N$ goes by the name of Fekete's lemma.
To prove it, pick integers $M$ and $P$, and note that we inductively have
\begin{equation} \label{eq:subadditivity_induction}
\frac{F^{(KM+P)}}{KM+P} \leq \frac{K F^{(M)}}{KM+P} + \frac{F^{(P)}}{KM+P}.
\end{equation}
Taking $K \rightarrow \infty$ gives
\begin{equation} \label{eq:subadditivity_intermediate_step}
\limsup_{K \rightarrow \infty} \frac{F^{(KM+P)}}{KM+P} \leq \frac{F^{(M)}}{M}.
\end{equation}
This holds for all $P \in \{0, 1, \cdots, M-1\}$, thus $\limsup_{N \rightarrow \infty} F^{(N)}/N \leq F^{(M)}/M$.
Taking the liminf as $M \rightarrow \infty$ then gives
\begin{equation} \label{eq:subadditivity_final_step}
\limsup_{N \rightarrow \infty} \frac{F^{(N)}}{N} \leq \liminf_{M \rightarrow \infty} \frac{F^{(M)}}{M},
\end{equation}
i.e., the two must be equal and the limit exists.

Having established that $f \equiv \lim_{N \rightarrow \infty} F^{(N)}/N$ exists, now turn to Eq.~\eqref{eq:free_energy_difference_expression_reproduced}.
Since $F^{(N)} \geq F^{(N+L)} - F^{(L)}$ for all $L$, we certainly have that $F^{(N)} \geq \limsup_{L \rightarrow \infty} [F^{(N+L)} - F^{(L)}]$.
Dividing by $N$ and taking $N \rightarrow \infty$ then gives
\begin{equation} \label{eq:free_energy_difference_expression_one_side}
f \geq \limsup_{N \rightarrow \infty} \limsup_{L \rightarrow \infty} \frac{F^{(N+L)} - F^{(L)}}{N}.
\end{equation}
At the same time, we have that for all $N$,
\begin{equation} \label{eq:free_energy_difference_expression_other_side_v1}
\begin{aligned}
f = \lim_{j \rightarrow \infty} \frac{F^{(jN)}}{jN} &= \lim_{j \rightarrow \infty} \frac{1}{j} \sum_{i=0}^{j-1} \frac{F^{((i+1)N)} - F^{(iN)}}{N} \\
&\leq \limsup_{j \rightarrow \infty} \frac{F^{((j+1)N)} - F^{(jN)}}{N} \leq \limsup_{L \rightarrow \infty} \frac{F^{(N+L)} - F^{(L)}}{N}.
\end{aligned}
\end{equation}
Taking $N \rightarrow \infty$ then gives
\begin{equation} \label{eq:free_energy_difference_expression_other_side_v2}
f \leq \liminf_{N \rightarrow \infty} \limsup_{L \rightarrow \infty} \frac{F^{(N+L)} - F^{(L)}}{N},
\end{equation}
and Eq.~\eqref{eq:free_energy_difference_expression_reproduced} follows.

Lastly, we needed analogues of these results in Sec.~\ref{subsec:generalized_Aizenman_Sims_Starr_min_max}, where we considered the partition function $Z^{(N)}(R)$ and corresponding free energy $F^{(N)}(R)$ of states restricted to have a certain value $R$ of the self-overlap (technically a matrix $R_{\tau \tau'}$).
We proved in Sec.~\ref{subsec:generalized_Aizenman_Sims_Starr_min_max} the following analogue of sub-additivity: $F^{(N,L)}(R, R) \leq F^{(N)}(R) + F^{(L)}(R)$, where $F^{(N,L)}(R, R)$ is the free energy of states in a size-$(N+L)$ system restricted to \textit{separately} have $\sigma \cdot \sigma = R$ and $\alpha \cdot \alpha = R$ --- recall that we divided the spins into $\{ \vec{\sigma}_i \}_{i=1}^N$ and $\{ \vec{\alpha}_j \}_{j=1}^L$, and defined
\begin{equation} \label{eq:fully_generic_dot_products_reproduced}
\sigma(\tau) \cdot \sigma(\tau') \equiv \frac{1}{N} \sum_i \sigma_i(\tau) \sigma_i(\tau'), \qquad \alpha(\tau) \cdot \alpha(\tau') \equiv \frac{1}{L} \sum_j \alpha_j(\tau) \alpha_j(\tau').
\end{equation}

In fact, sub-additivity of the sequence $F^{(N)}(R)$ follows from this result simply by observing that the set of states with $\sigma \cdot \sigma = \alpha \cdot \alpha = R$ is a subset of the states with total self-overlap $R$ (note that the total self-overlap can be written $(N \sigma \cdot \sigma + L \alpha \cdot \alpha)/(N+L)$).
Thus $Z^{(N+L)}(R) \geq Z^{(N,L)}(R, R)$ since the sum that is $Z^{(N+L)}(R)$ includes every term of $Z^{(N,L)}(R, R)$, and $F^{(N+L)}(R) \leq F^{(N,L)}(R, R)$.
Fekete's lemma then proves that $f(R) \equiv \lim_{N \rightarrow \infty} F^{(N)}(R)/N$ exists, and a straightforward generalization of Eqs.~\eqref{eq:free_energy_difference_expression_one_side} through~\eqref{eq:free_energy_difference_expression_other_side_v2} gives
\begin{equation} \label{eq:generalized_free_energy_difference_expression_derivation}
\begin{aligned}
\limsup_{N \rightarrow \infty} \limsup_{L \rightarrow \infty} \frac{F^{(N,L)}(R, R) - F^{(L)}(R)}{N} \leq f(R) &\leq \liminf_{N \rightarrow \infty} \limsup_{L \rightarrow \infty} \frac{F^{(N+L)}(R) - F^{(L)}(R)}{N} \\
&\qquad \leq \liminf_{N \rightarrow \infty} \limsup_{L \rightarrow \infty} \frac{F^{(N,L)}(R, R) - F^{(L)}(R)}{N}.
\end{aligned}
\end{equation}
Thus
\begin{equation} \label{eq:generalized_free_energy_difference_expression_reproduced}
f(R) = \lim_{N \rightarrow \infty} \limsup_{L \rightarrow \infty} \frac{F^{(N,L)}(R, R) - F^{(L)}(R)}{N},
\end{equation}
which is Eq.~\eqref{eq:generalized_free_energy_difference_expression} from the main text.

\subsection*{Acknowledgements}

It is a pleasure to thank L. Foini for valuable and informative discussions.
C.L.B. was supported by the AFOSR, AFOSR MURI, DoE ASCR Quantum Testbed Pathfinder program (award No.~DE-SC0019040), DoE ASCR Accelerated Research in Quantum Computing program (award No.~DE-SC0020312), DoE QSA, NSF QLCI (award No.~OMA-2120757), NSF PFCQC program, ARO MURI, and DARPA SAVaNT ADVENT.
B.S. was supported by the DoE (award No.~DE-SC0009986).

\subsection*{Conflicts of interest}

The authors have no conflicts of interest to declare.

\subsection*{Data availability}

No datasets were generated during this work, but code and/or further details of calculations are available from the corresponding author upon request.

\bibliography{biblio}

\end{document}